\documentclass[a4paper,fleqn,usenatbib]{mnras}

\usepackage[T1]{fontenc}
\usepackage{ae,aecompl}

\usepackage{graphicx}	% Including figure files
\usepackage{amsmath}	% Advanced maths commands
\usepackage{amssymb}	% Extra maths symbols
\usepackage{mathtools}	% for multiline equations
\title[Pulsar searching with the Fast Folding Algorithm]{An investigation of pulsar searching techniques with the Fast Folding Algorithm}

\author[A. D. Cameron et al.]{
A. D. Cameron$^{1}$,\thanks{E-mail: acameron@mpifr-bonn.mpg.de}
E. D. Barr$^{1,2}$,
D. J. Champion$^{1}$,
M. Kramer$^{1}$,
W. W. Zhu$^{1}$
\\
$^{1}$Max-Planck Institut f{\"u}r Radioastronomie, Auf dem H{\"u}gel 69, D-53121 Bonn, Germany\\
$^{2}$Centre for Astrophysics and Supercomputing, Swinburne University of Technology, Mail H39, PO Box 218, \\Hawthorn, VIC 3122, Australia\\
}

\date{Accepted XXX. Received YYY; in original form ZZZ}

\pubyear{2017}

\begin{document}
\label{firstpage}
\pagerange{\pageref{firstpage}--\pageref{lastpage}}
\maketitle

\begin{abstract}
Here we present an in-depth study of the behaviour of the Fast Folding Algorithm, an alternative pulsar searching technique to the Fast Fourier Transform. Weaknesses in the Fast Fourier Transform, including a susceptibility to red noise, leave it insensitive to pulsars with long rotational periods ($P > 1\text{ s}$). This sensitivity gap has the potential to bias our understanding of the period distribution of the pulsar population. The Fast Folding Algorithm, a time-domain based pulsar searching technique, has the potential to overcome some of these biases. Modern distributed-computing frameworks now allow for the application of this algorithm to all-sky blind pulsar surveys for the first time. However, many aspects of the behaviour of this search technique remain poorly understood, including its responsiveness to variations in pulse shape and the presence of red noise. Using a custom CPU-based implementation of the Fast Folding Algorithm, \texttt{ffancy}, we have conducted an in-depth study into the behaviour of the Fast Folding Algorithm in both an ideal, white noise regime as well as a trial on observational data from the HTRU-S Low Latitude pulsar survey, including a comparison to the behaviour of the Fast Fourier Transform. We are able to both confirm and expand upon earlier studies that demonstrate the ability of the Fast Folding Algorithm to outperform the Fast Fourier Transform under ideal white noise conditions, and demonstrate a significant improvement in sensitivity to long-period pulsars in real observational data through the use of the Fast Folding Algorithm. 
\end{abstract}


%%%%%%%%%%%%%%%%%%%%%%%%%%%%%%%%%%%%%%%%%%%%%%%%%%

%%%%%%%%%%%%%%%%% BODY OF PAPER %%%%%%%%%%%%%%%%%%

\section{Introduction}
\label{Intro}

The Fast Fourier Transform (FFT) is one of the most efficient and widely used techniques in pulsar searching \citep{lk05}. However, despite the success of this algorithm in many large-scale pulsar search campaigns, it is not without its limitations. In particular, FFT-based search pipelines remain vulnerable to the presence of red noise, which can significantly limit their sensitivity to long-period pulsars (defined in this paper as pulsars with a rotational period $P > 1\text{ s}$). A notable demonstration of these limitations was made by \cite{lbh15} as part of an investigation into possible discrepancies between the theoretical and actual sensitivities of the PALFA survey \citep{cfl+06}. Through a process of attempting to recover synthetic pulsar profiles that had been injected into real observational data, it was demonstrated that significant losses in sensitivity occurred for pulsars with periods longer than $P \simeq 100 \text{ms}$. This degradation in sensitivity was seen to increase with increasing rotational period, resulting in the loss of approximately $35\%$ of the pulsars which PALFA may otherwise have detected. This effect has the potential to introduce a strong selection bias in our current searching pipelines, skewing our knowledge of the underlying source distribution. 

An alternative approach to pulsar searching is to instead search directly in the time domain by folding an observational time series and examining the resulting folded profiles. The Fast Folding Algorithm \citep[FFA,][]{sta69} is a computationally efficient means of folding multiple trial periods. This efficiency is achieved through the removing of redundant computational steps in the folding process and by storing the results of these steps for later recall as needed. 

The FFA presents a number of key advantages in searching for pulsars in the long-period regime. For example, the operational speed of the FFA increases when folding over longer periods, as the number of addition operations required during the folding process is of the order $O(N\log_2(N/P_0))$, where $N$ is the number of samples in the original dataset and $P_0$ is the base trial period being folded over (in units of samples). Thus for fixed $N$, the number of addition operations decreases as $P_0$ increases, giving the FFA a performance boost at longer trial periods. Furthermore, although it is not immune to red noise, the FFA is able to produce a result that is fully coherent in phase, unlike the FFT which is normally only able to incoherently sum a limited number of harmonics \citep{lbh15}, typically 16 or 32 harmonics depending upon the specific FFT implementation. This results in some of the pulsar's Fourier power remaining at unsummed harmonic frequencies, reducing its overall sensitivity. Therefore, the FFA may present a viable solution for recovering some of the sensitivity previously lost in the long-period regime.

Detailed research into the properties of the FFA was previously conducted by \cite{kml09}. Through the use of a custom implementation of the FFA, \texttt{ffasearch}\footnote{http://astro.wvu.edu/projects/xdins}, a comparison of the performance of both the FFT and FFA was produced. This comparison was conducted both from an analytical standpoint, comparing the theoretical sensitivity of each technique, as well as from an experimental standpoint, through a simulation performed over a range of artificial pulsars created using the \texttt{SIGPROC} program \texttt{fake}. These artificial pulsars covered a range of periods $P$ between $2{\rm s}$ and $14{\rm s}$ and duty cycles ($\delta$) between $0.1\%-5\%$, with the total per-pulse energy kept consistent for each pulsar trial. The results of both the theoretical and experimental comparisons clearly demonstrate that the in the narrow duty cycle regime ($\delta < 1\%$), the FFA displays a much greater sensitivity than the FFT to pulsars with periods longer than $\sim2{\rm s}$, and that this remains the case for larger values of $\delta$ at even longer periods. This further builds the case for the FFA as a useful searching tool in the long-period pulsar regime.

However, unlike the FFT, to date the FFA has seen relatively little use in large-scale pulsar searches. Notable examples of its use include the discovery of J2018+2839 \citep{csl68}, the investigation of quiescent RRAT radio emission through the folding of time series around the known period of existing RRATs \citep{ld14}, the discovery of a $7.7\text{ s}$ pulsar in the Parkes Multibeam Pulsar Survey \citep{fsk+04, lfl+06} and a search for pulsed radio emission from the X-ray pulsar XTE J0103-728 \citep{cld09}. This limited usage is primarily due to the fact that despite its inherent efficiency over other, simpler folding techniques, it has remained prohibitively computationally expensive. However, as computational power continues to increase, these concerns are gradually becoming less significant. Also, the FFA is highly parallelisable, making it suitable for application to a parallel processing framework of general-purpose, many-core accelerators, allowing for further increases in performance.

Of course, the FFA is not without its own set of challenges to overcome. Chief amongst these is that unlike the FFT, which produces a power spectrum and the interpretation of which is well understood, the FFA produces a series of folded pulse profiles. Depending on the precise configuration of the FFA, these folded profiles can number in excess of several million per dispersion trial. This necessitates the development of algorithms which can evaluate the likelihood of a given folded profile representing the detection of a pulsar via a representative score or algorithm. These algorithms must themselves also be computationally efficient, as they must be applied to every single folded profile produced by the FFA, and their behaviour must be well understood in terms of their responsiveness to different pulse shapes, duty cycles, and other related parameters. The grouping of the resulting scores into appropriate lists of potential pulsar candidates must also be dealt with. Finally, the response of the FFA to the presence of red noise as well as other forms of RFI, and the degree to which this influence can be mitigated, remains to be investigated.

This challenge of pulse profile evaluation is in some ways not dissimilar to the problems facing another time-domain pulsar searching technique, the single-pulse search. This technique attempts to detect the bright individual pulses of pulsars and other radio transients (such as Rotating Radio Transients (RRATs) and Fast Radio Bursts (FRBs)) through the analysis of de-dispersed time series. The basic modern technique, as presented by \cite{cm03}, involves the identification of significant outlying data points in a given time series, with matched-filtering used in order to tackle single pulses of varying width, with the majority of continuing single-pulse search efforts \citep[for recent work see e.g.,][]{htru3, dgm16, aa16} building upon this foundation. Elements of these existing time-domain based approaches may be able to be utilised in developing profile evaluation algorithms for the FFA.

In this paper, we present the results of a new investigation into the behaviour of the FFA, building upon the work conducted by \cite{kml09}. This includes an analysis of the performance and robustness of multiple algorithms in evaluating the results of our own implementation of the FFA against a theoretical ideal FFA implementation and both an ideal and actual FFT implementation. Our investigation also includes the response of both the FFA and FFT to red noise contamination. The focus of this paper is on the computational correctness, robustness and response of the algorithms used, rather than the optimisation of their implementation. The question of designing and producing a computationally efficient implementation of the FFA and its associated evaluation algorithms (likely with the use of parallel processing) will be discussed in a future publication.

The structure of this paper is as follows: Section \ref{sec:Background} outlines the mathematical theory of the FFA and our chosen algorithms, and describes the implementation of the FFA used in this study. Section \ref{MetricTesting} then explores the behaviour and response of each profile evaluation algorithm under a variety of controlled conditions. Section \ref{sec:realtrial} presents a demonstration of the FFA using real-world observational datasets and explores responses of the FFA to the presence of red noise and other RFI. A brief discussion and conclusion then follow in Sections \ref{sec:Discussion} and \ref{sec:Conclusion} respectively.

\section{Mathematical Background \& Implementation}\label{sec:Background}

As the majority of the results presented in this paper rely upon the specific details of each of the algorithms chosen for testing, as well as their specific implementation, this section will first outline the necessary background information.

\subsection{The Fast Folding Algorithm (FFA)}

The full details and background of the FFA are laid out in \cite{sta69}, which also includes a schematic of the algorithm's operation. Supplementary information is also available in \cite{lss69}, \cite{bc69} and \cite{lk05}. Provided here is a brief outline of the fundamental aspects and behaviour of the FFA.

Consider a dataset which has a length of $N$ samples (e.g., in the case of pulsar analysis, a dedispersed timeseries). A single execution of the FFA conducts a periodicity search over this dataset using a base period of $P_0$ samples (which can be converted into units of time by multiplying by the relevant sampling interval $\Delta t$). The FFA does this by breaking the dataset down into segments that are $P_0$ samples long. Simply adding these segments together would result in a single folded profile with a period of $P_0$ samples, however, by applying relative offsets to each of these segments before adding them together, and by storing redundant operational results, the FFA is able to efficiently fold $N/P_0$ folded profiles in a single execution. These folded profiles correspond to a range of trial periods between the base period $P_0$ and $P_0 + 1$ inclusive, in increments of \begin{equation}\Delta P=\frac{P_0}{N - P_0}\end{equation} samples. Therefore, the trial period $P_i$ for any given folded profile can be given by \begin{equation}P_{i}=(P_{0} + \frac{P_{0}}{N-P_{0}}i)\end{equation} where $0 \leq i < (N/P_{0})$ and $i$ is an integer. By conducting multiple FFA executions for differing values of $P_0$, a large range of periods can be searched.

% A large chunk of this paragraph might be better off in the software section
The most efficient implementation of the FFA requires that the ratio $N/P_{0}$ be equal to some power of two, that is, $\log_{2}(N/P_{0})$ must be an integer. In order to test a wide range of arbitrary periods which do not necessarily meet this requirement, it is therefore necessary to either pad or trim the original dataset of $N$ samples (producing some new dataset of length $N_{*}$ samples) such that the correct relationship is maintained (i.e. $\log_{2}(N_{*}/P_{0})$ is an integer). In order to preserve as much of the original dataset as possible, we have chosen in our implementation to only ever increase the value of $N$ such that $N_{*} \geq N$, using zero values as extra data elements so as not to introduce any additional noise into the dataset. As this transition from observational data to zeroes could potentially cause an artificial jump in the baseline of the folded profiles, a small amount of observational data at the end of the observation is also set to zero such that the length of the 
remaining original dataset is equal to an integer multiple of $P_0$ for each execution of the FFA.

\subsection{Profile Evaluation Algorithms}

In order to evaluate the folded profiles produced by the FFA in a regular and efficient manner, we have developed multiple profile evaluation algorithms to be applied to each individual folded profile. These algorithms are able to break down a folded profile into a simple numerical score, with the aim that this score should in some way positively correlate to a profile's probability of representing a pulsar detection. These algorithm scores can then be plotted as a function of period to form a periodogram, and can be further analysed in the process of candidate selection. As previously mentioned, a critical property of any algorithm used for profile evaluation is its speed, as it will need to run over every single folded profile produced by the FFA. Therefore, while potentially ``ideal'' algorithms can be imagined which can account for every single pulse profile scenario, some simplifications will be required in the algorithms chosen for implementation in order to retain a lower computational expense.

Although many algorithms were initially tested during the early research stages of this project, only the following two were selected for rigorous testing, due to the significantly higher sensitivities and robustness demonstrated by each during preliminary testing, as well as their ability to compute within a reasonable CPU time. Both algorithms work under the assumption that, aside from the presence of any pulsar in the data, the profiles to be evaluated contain normally distributed noise. Each algorithm is designed to produce an approximation of the time-domain profile signal-to-noise ratio (S/N) as its ``score'', using one of two different approaches. It should be noted that these measures of S/N are produced on a profile-by-profile basis, and that they do not represent a measure of the overall significance of a detected signal (a quantity which is dependent on the total number of evaluated profiles). The robustness of each algorithm to variations in the pulse profile is further explored in Section \ref{MetricTesting}.

\subsubsection{Algorithm 1: A boxcar matched-filter with Median Absolute Deviation (MAD) normalisation}
\label{Metric6} 

Algorithm 1 employs the use of a series of scrolling boxcar matched-filters of the same style as described by \cite{cm03} in an attempt to capture all of the pulse profile power within a single bin, while simultaneously capturing as little off-pulse noise as possible. Before this step however, the algorithm attempts to normalise each folded profile (reducing the off-pulse baseline of the profile to 0 and reducing the profile standard deviation $\sigma$ to 1) so as to assist in later calculations. This normalisation is performed using a statistical quantity known as the Median Absolute Deviation (MAD), which is defined for a finite set of values $x_1, x_2, \dots x_n$ as
\begin{equation}
  \text{MAD} = \text{median}(|x_i - \text{median}(\vec{x})|).                                                                                                                                                                                                                                                                                                                                                                                                                                                                                                                                                                                                                                                                                                                                                                                                                                                                                                                   \end{equation}  
                                                                                                                                                                                                                                                                                                                                                                                                                                                         That is, it represents the median of the absolute deviations of each value from the median of the entire dataset. For large, normally distributed datasets, MAD is related to the standard deviation $\sigma$ by
\begin{equation}
\sigma = K \cdot \text{MAD}                                                                                                                                                                                                                                                                                                                                                                                                                                                                                                                                                                                                                                                                                                                                                                                                                                                                                                                                                                                                                                            
                                \end{equation} 
where $K \approx 1.4826$. However, unlike $\sigma$, MAD is a much more robust statistic as it remains more resilient to the presence of outliers in the data, due to the fact that it is based on the median of the dataset rather than the mean. Therefore, so long as the pulse contained in a folded profile is narrow (an appropriate approximation in the case of the majority of long-period pulsars \citep{tml93}, although exceptions do exist), MAD can be used to quickly normalise a folded profile as if it contained no pulse, as the ``outlying'' data points of the pulse are, to a good approximation, ignored. MAD normalisation is hence performed by first subtracting the median from the folded profile (so as to reduce the baseline to 0), then by dividing all values by $\text{MAD} \cdot K$ (so as to give unity variance).

Once a folded profile has been normalised in this manner, the value of any individual bin can be taken as its S/N. However, as most pulse profiles are wider than a single bin, successively larger boxcar matched-filters (up to some defined maximum filter width) are applied to the profile at each bin position in order to rebin the data at different pulse widths. In our chosen implementation, we use matched-filters at sizes of $2^n$ samples, where $n \geq 0$ is an integer, with the largest filter chosen as the first filter size to exceed $\delta=20\%$. Normalisation is maintained by dividing each newly rebinned value by the square root of the width of the applied matched-filter. If a pulse is present in the profile, it should produce a maximum signal at the filter closest to its true pulse width, capturing as much pulse and as little noise as possible. The maximum value from all of these rebinned profiles is taken as the algorithm score, $A_1$, for 
that particular profile. An example of the response of Algorithm 1 may be found in Figure \ref{fig:MetricDemo_figure}.

\subsubsection{Algorithm 2: A boxcar matched-filter with an off-pulse window}
\begin{figure}
	\includegraphics[width=\columnwidth]{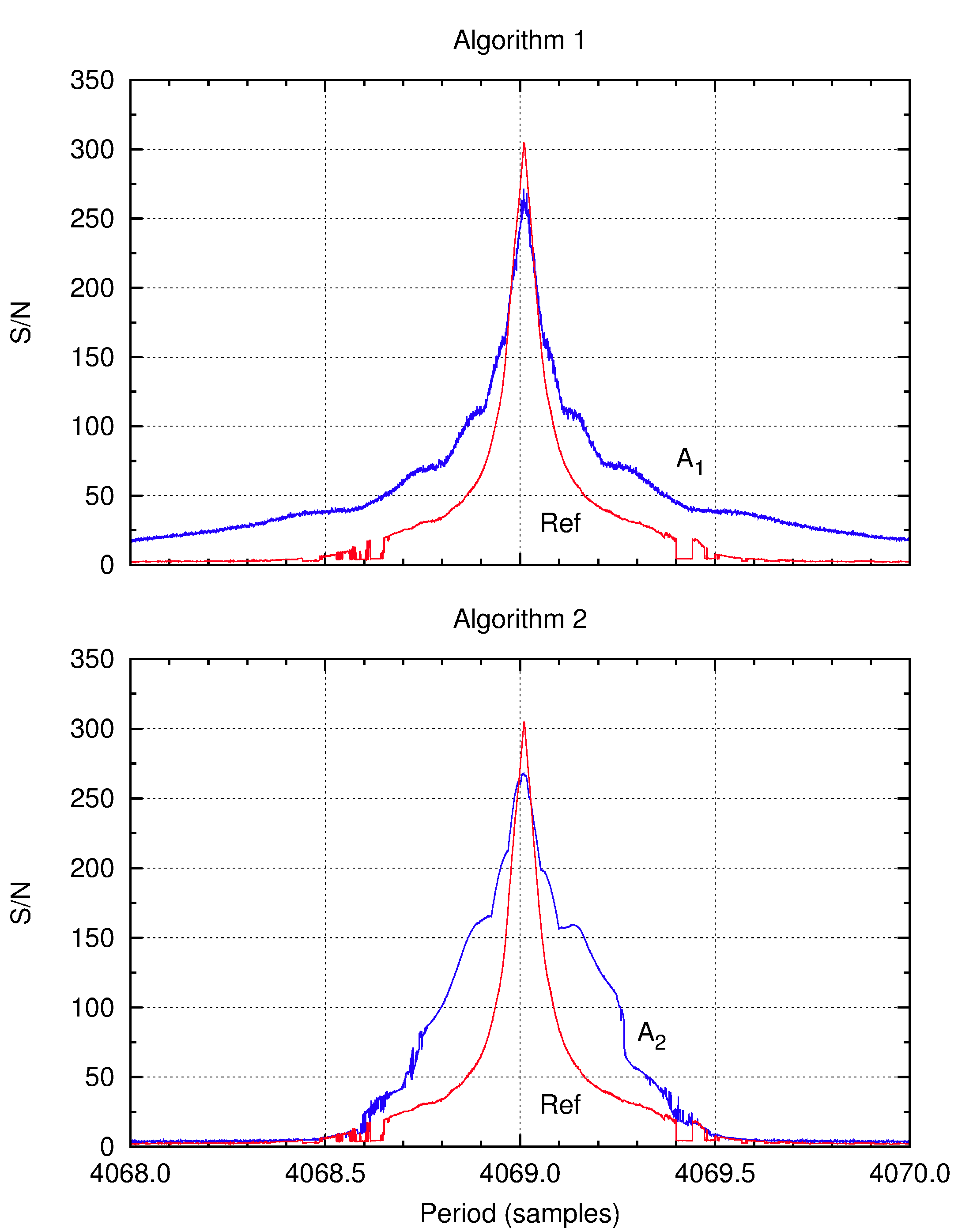}
    \caption{Zoomed-in view of a demonstration periodogram response for Algorithms 1 \& 2 for a simulated pulsar with a top-hat pulse profile against a background of white noise. The pulsar has a period of approximately 4069 samples and a duty cycle $\delta=1.0\%$. The ``bumps'' in each curve indicate points where a new matched-filter gives a stronger algorithm response. The reference curve (marked ``Ref'') indicates the response of a secondary version of Algorithm 2 which uses a pre-specified matched-filter in its evaluation. This filter was selected to optimally match the precise width of the top-hat pulsar, which in this case is 40 samples wide.}
    \label{fig:MetricDemo_figure}
\end{figure}

Algorithm 2 essentially borrows its technique from the work performed by \cite{kml09}. The matched-filtering technique as described by \cite{cm03} is also applied here, using the same set of filter sizes as Algorithm 1, but unlike Algorithm 1 the matched-filtering is applied first, without applying any normalisation beforehand. After the application of the matched-filter, the maximum value in the profile, $I_{max}$, is then identified. An exclusion window corresponding to $20\%$ of the profile width is then centered over the position of $I_{max}$ on the profile, removing the ``on-pulse'' data values from further consideration. The average value and standard deviation of the profile, $I_{av}$ and $\sigma$ are then calculated based on the remaining $80\%$ of the profile, the ``off-pulse'' component. The S/N $A$ is then given by
\begin{equation}
 A = (I_{max} - I_{av})/\sigma
\end{equation}
The maximum value of $A$ over all of the matched-filters for a particular folded profile is then selected as the final algorithm score, $A_2$. An example of the response of Algorithm 2 may be found in Figure \ref{fig:MetricDemo_figure}.

\subsection{Software}

\subsubsection{FFAncy}\label{subsubsec:ffancy software}

A core part of the work of this project involved the development and testing of our own custom CPU-based implementation of the FFA, dubbed \texttt{ffancy}\footnote{https://github.com/adcameron/ffancy}. Written in C, \texttt{ffancy} was designed as a testbed program to allow for a full investigation of the FFA's behaviour. In its current version, it is capable of either generating its own simple internal datasets or reading an externally generated dedispersed time series, before conducting the FFA over a period search range specified in units of samples. \texttt{ffancy} is presently capable of reading both 8-bit unsigned-integer time series as produced by the \texttt{SIGPROC}\footnote{https://github.com/SixByNine/sigproc} pulsar software package, or 32-bit float time series as produced by the \texttt{PRESTO}\footnote{https://github.com/scottransom/presto} \citep{ransom01} pulsar software package. In addition to the standard periodogram output of algorithm scores against trial period values, \texttt{ffancy} is also capable of producing multiple output streams for use in 
testing, including the full suite of folded profiles produced by a full-scale FFA execution as well as dereddened copies of the original input time series. \texttt{ffancy} exports a simple command line interface that allows for the selection of period ranges, algorithms, etc. at runtime. It also employs a modular, extensible design that simplifies the addition and testing of new algorithms.

While the implementation of the FFA at the program's core is essentially the same as the original algorithm proposed by \cite{sta69}, some additional features have also been added in order to optimise performance during testing. These include the ability to pre-downsample the data before execution of the FFA in order to increase the speed of a search by reducing the number of samples, as well as an automatic downsampling routine whereby a search conducted with a lowest trial period of $P$ samples will automatically downsample the dataset by a factor of 2 every time a period of $2^n\cdot P$ is reached, where $n$ is an integer. A dereddening option (which employs a dynamic running median filter) can also be selected at runtime. By default, the size of the filtering window used in the dereddening is initially set to twice the maximum period to be searched or four times the lowest period to be searched, whichever is smaller. Alternatively, the user is able to manually specify their own window size at 
runtime. This dereddening scheme is dynamic in that the dereddening window doubles in size relative to the initial time series concurrently with each downsampling during the execution of the FFA search, such that the window size is always appropriate to the range of periods currently being searched.

\subsubsection{Progeny, Prostat \& Metrictester}

As much of the testing conducted as part of this paper involves evaluating the performance characteristics of the algorithms themselves, a separate simple program termed \texttt{progeny} was developed in order to produce simulated folded profiles as would be output by \texttt{ffancy} but without the need for a full FFA execution. \texttt{progeny} is able to generate custom profiles with or without white noise, and allows the user to choose the baseline, noise level, and number of bins in the profile, as well as the height, width and position of any seeded pulse. It has the ability to seed either top-hat pulses or Gaussian pulses, with the width of such a Gaussian pulse determined as its full width at half maximum (FWHM). The user can specify as many or as few pulse components as required, allowing for the generation of complex, multi-component pulse profiles. Finally, the program is also able to simulate the effects of scattering by convolving any seeded pulse shape with a one-sided exponential decay function. Each algorithm can be used to evaluate a \texttt{progeny} format profile using the additional application \texttt{metrictester}, and a further application \texttt{prostat} can be used to provide a simple statistical breakdown of the contents of each \texttt{progeny} profile.

\subsubsection{ffa2best}

In order to refine the large number of algorithm scores produced as part of each FFA execution down to a manageable list of candidates, an additional program termed \texttt{ffa2best} was also developed. This program is designed to accept the periodogram output produced by \texttt{ffancy} and convert it into a candidate list formatted in the style of the \texttt{SIGPROC} program \texttt{best}. The process of candidate selection is complicated by the fact that each ``peak'' in a periodogram has some finite width, with the recorded algorithm S/N rising near the optimal period of the candidate (see Figure \ref{fig:MetricDemo_figure}). The basic functionality of \texttt{ffa2best} allows the seperation of candidates in the periodogram using a series of 3 thresholds, all of which can be specified at runtime. These thresholds determine the baseline S/N below which candidates will not be recorded, as well as a fractional and absolute threshold by which the S/N has to fall from the previous candidate before a new candidate can be recorded. More advanced functionality allows for the grouping of nearby candidates based upon the expected response of a strong pulsar whose profile is smeared due to an incorrect period (as determined from the results of the tests outlined in Section \ref{MetricTesting}), and for the matching and removal of harmonically-related candidates. This harmonic matching is performed using each of the prime numbers up to a user-specified maximum. For each prime $x$, a given candidate is both multiplied and divided by each fraction $n/x$ where $0<n<x$ is an integer, so as to account for both higher and lower harmonic frequencies. The use of primes prevents the repetition of redundant fractions. Two candidates are treated as harmonically related if a match is obtained using this procedure to within a specified tolerance factor, with the weaker candidate being removed from the final candidate list.

\subsection{Testing for correctness}
\label{KondratievCorrectness}
In addition to the standard testing of code required as part of the software development process, a critical testing milestone for \texttt{ffancy} was for it to demonstrate an ability to confirm the results of the simulation performed by \cite{kml09}. Our simulation was conducted by generating artificial pulsars using the \texttt{SIGPROC} program \texttt{fake}, with periods $P$ between $2s$ and $14s$ inclusive (with an increment of $0.5s$) and with duty cycles $\delta$ covering $0.1\%-1\%$ (increments of $0.1\%$) and $2\%-5\%$ inclusive (increments of $1\%$). Each of the resulting 350 combinations of $P$ and $\delta$ was generated and tested 30 times, so as to mitigate the influence of any statistical anomalies which may arise in a single dataset, resulting in a total of 10500 trial pulsars. The total per-pulse energy, given by \begin{equation}\label{E=wS}
E_p = wS\end{equation} 
for the top-hat pulses as seeded by \texttt{fake} (where $S$ is the height of the pulse and $w=\delta P$ is the width of the pulse), was kept consistent between each pulsar trial. The sampling time set to $\Delta t=491.52\mu s$ and observation time set to $t_{obs} = 3600 s$ so as to match the original simulation parameters from \cite{kml09}. 

Each pulsar was processed through both the FFA and the FFT, with the resulting S/N of the detected signal at the expected period extracted and averaged over the 30 trials for each combination of $P$ and $\delta$. The FFA analysis was conducted using \texttt{ffancy}, which conducted a narrowly targeted period search on each trial pulsar at full time resolution. The chosen algorithm was a secondary version of Algorithm 2 which employs a pre-specified matched-filter chosen to match the width of the seeded pulse, rather than the blind power-of-2 filter normally used by Algorithm 2. This most closely matches the algorithm used in the original simulation by \cite{kml09}, and should theoretically result in the maximal S/N. The FFT analysis was conducted using the \texttt{SIGPROC} program \texttt{seek}, with the representative S/N detection being taken from the candidate with the maximum S/N in the resulting candidate lists, regardless of the selected harmonic. It should be noted that due to software limitations 
with \texttt{seek}, our simulation employed only 16 harmonic sums compared to the 32 used in the original simulation. This resulted in a slight loss in sensitivity across all test cases.

In order to provide an objective baseline of sensitivity, both the FFA and the FFT were also compared against their theoretically expected S/Ns for each combination of $P$ and $\delta$. The analytical expressions for both the FFA and FFT were derived as part of \cite{kml09} and will not be re-derived here. In the case of the FFA, assuming a top-hat pulse shape and the application of a matched-filter equal in size to the seeded pulse width $w$, for a pulsar with a period of $P$, duty cycle of $\delta$ and a total per-pulse energy of $E_p$ recorded over a dataset of length $N$, the theoretical maximum S/N of the detected pulsar signal is given by\begin{equation}\label{FFA_sensitivity} 
\text{S/N}_\text{ffa}=\frac{E_p\sqrt{N}}{P\sqrt{\delta}} .                                                                                                                                                                                                                                                                                                                                                                                                                                                                                                                                                                                                        
                                                                                                                                                                                                                                                                                                                                                                                                                                                                                                                                                                                                 \end{equation} 
                                                                                                                                                                                                                                                                                                                                                                                                                                                                                                                                                                                                 
                                                                                                                                                                                                                                                                                                                                                                                                                                                                                                                                                                                                 In the case of the FFT, for a pulsar with a duty cycle of $\delta$, a pulse height of $S$, recorded over a number of samples $N$ and summed harmonically over $H$ harmonics, the theoretical S/N of the detected pulsar signal is given by
                                                                                                                                                                                                                                                                                                                                                                                                                                                                                                                                                                                                 \begin{equation}
                                                                                                                                                                                                                                                                                                                                                                                                                                                                                                                                                                                              \label{eqn:FFT_sensitivity}
                                                                                                                                                                                                                                                                                                                                                                                                                                                                                                                                                                                              \begin{aligned}
                                                                                                                                                                                                                                                                                                                                                                                                                                                                                                                                                                                             \text{S/N}_\text{fft}=\sqrt{\frac{\pi}{H(4-\pi)}}\times \\ 
                                                                                                                                                                                                                                                                                                                                                                                                                                                                                                                                                                                             \sum_{n=1}^{H}[L_{1/2}^0(-N[S(1-\delta)\text{sinc}[\pi n(1-\delta)]]^2)-1]                                                                                                                                                                                                                                                                                                                                        
                                                                                                                                                                                                                                                                                                                                                                                                                                                                                                                                                                                                                                                                                                                                                                                                                     \end{aligned}
                                                                                                                                                                                                                                                                                                                                                                                                                                                                                                                                                                                                                                                                                                                                                                                                                     \end{equation} where $L_{1/2}^{0}(x)$ is the generalised Laguerre polynomial $L_{n}^{\alpha}(x)$ where $n=1/2$ and $\alpha=0$. It should be noted that although both of these equations provide the expected S/N for a given pulsar signal, Equation \ref{FFA_sensitivity} represents a time domain S/N measure while Equation \ref{eqn:FFT_sensitivity} represents a Fourier domain S/N measure.

\begin{figure*}
	\includegraphics[width=\textwidth]{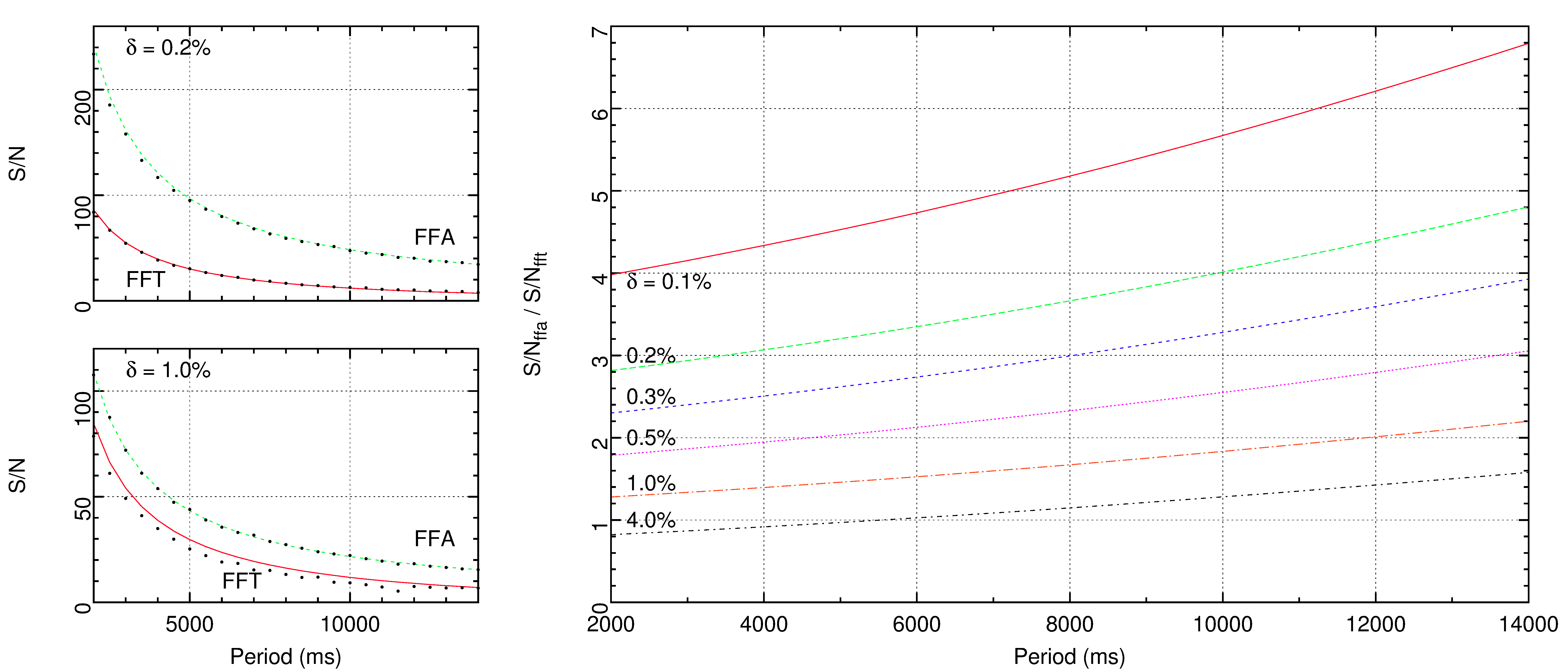}
    \caption{Simulation results showing confirmation of the work carried out as part of \protect\cite{kml09}, namely Figure 8. Left: S/N of both the FFA and FFT at two different duty cycles. The analytical predictions are shown by continuous coloured lines while the experimental results from the simulation are shown as points. Right: the ratio of $\text{S/N}_\text{FFA}$ to $\text{S/N}_\text{FFT}$ as a function of period and duty cycle. These results are taken from the analytical expressions in equations \protect\ref{FFA_sensitivity} and \protect\ref{eqn:FFT_sensitivity}.}
    \label{fig:Kondratiev_figure}
\end{figure*}

The results of the simulation can be seen in Figure \ref{fig:Kondratiev_figure}, which has been produced to emulate the original Figure 8 from \cite{kml09}. A comparison between these two figures demonstrates similar trends in both simulations, with the advantages of the FFA in the narrow duty cycle range extremely clear. The loss in sensitivity of the FFT due to the choice to only sum 16 harmonics is most clearly noticed in the case of $\delta = 1.0\%$, with the previous overlap in sensitivity at lower periods now absent. The ratio comparison of the two analytical curves also shows an increased favorability towards the FFA over the FFT in all but a few cases of shorter period pulsars with duty cycles of at least a few percent. Meanwhile, both example duty cycle plots show strong agreement between the theoretical predictions of the analytical curves described in Equations \ref{FFA_sensitivity} and \ref{eqn:FFT_sensitivity} and the experimental data. In summary, the results of the simulation clearly demonstrate both the ability of \texttt{ffancy} to confirm the results of the previous FFA study, and once again demonstrate the broad advantages of the FFA over the FFT in the long-period pulsar regime.

\section{Algorithm Testing}
\label{MetricTesting} 

As a first step towards understanding the behaviour of the FFA, it is necessary to conduct a full investigation of the behaviour of the algorithms chosen to evaluate its output. It is these algorithms that produce the fundamental data required to evaluate the presence of any potential pulsar in an observation. Therefore, any shortcomings in these algorithms in their ability to detect particular types or regimes of pulsar profiles must be understood. This section will outline the results of a series of tests intended to thoroughly probe the response patterns of both of the algorithms chosen for implementation in our study.

\subsection{Duty cycle, pulse height and pulse energy}\label{subsec: DHE} 

In the case of a simple, single-component, symmetric pulse, the three fundamental parameters of this pulse that can be modified are its width $w$, height $S$ and total energy or fluence $E$ (the area under the pulse curve). In the case of a top-hat pulse, these three parameters are linked simply by Equation \ref{E=wS}, where for a single folded profile the period $P$ is represented by the profile's length. Such a top-hat pulse results in the sensitivity equation given in Equation \ref{FFA_sensitivity}. Given that we are dealing in this case with a single profile, $P=N$ and this can be rewritten as\begin{equation}\label{Single pulse FFA sensitivity - DCE} 
\text{S/N}_\text{ffa}=\frac{E_p}{\sqrt{P\delta}}    .                                                                                                                                                                                                                                                                                                                                                                                                                                                                                                                                                                                                                                                                       
                                                                                                                                                                                                                                                                                                                                                                                                                                                                                                                                                                                                                                                                           \end{equation} 
                                                                                                                                                                                                                                                                                                                                                                                                                                                                                                                                                                                                                                                                           Using Equation \ref{E=wS}, this can also be rewritten as\begin{equation}\label{Single pulse FFA sensitivity - DCH} 
\text{S/N}_\text{ffa}=S\sqrt{\delta P}                 
                                                                                                                                                                                                                                                                                                                                                                                                                                                                                                                                                                                                                                                                                                                                                                 \end{equation}   and                                                                                                                                                                                                                                                                           
                                                                                                                                                                                                                                                                                                                                                                                                                                                                            \begin{equation}\label{Single pulse FFA sensitivity - EH} 
\text{S/N}_\text{ffa} = \sqrt{SE_p}    .                                                                                                                                                                                                                                                                                                                                                                                                                                                                                                                                                                                                                                                                                                                                                     
                                                                                                                                                                                                                                                                                                                                                                                                                                                                                                                                                                                                                                                                                                                                                                \end{equation}
                                                                                                                                                                                                                                                                                                                                                                                                                                                                                                                                                                                                                                                                                                                                                                 Based upon these expressions, we can predict the theoretical behaviour of each algorithm as we vary two out of three specified parameters, and compare the experimental behaviour of each algorithm to this prediction. Using \texttt{progeny}, a series of simulated top-hat profiles were 
created in order to explore this parameter space. The number of bins in each profile was kept at 1024, and each profile test was repeated 30 times using newly generated noise in each repetition, with the average taken as the overall result.

\subsubsection{Duty cycle vs. pulse height}\label{subsec: DCH} 

\begin{figure} 
 \includegraphics[width=\columnwidth]{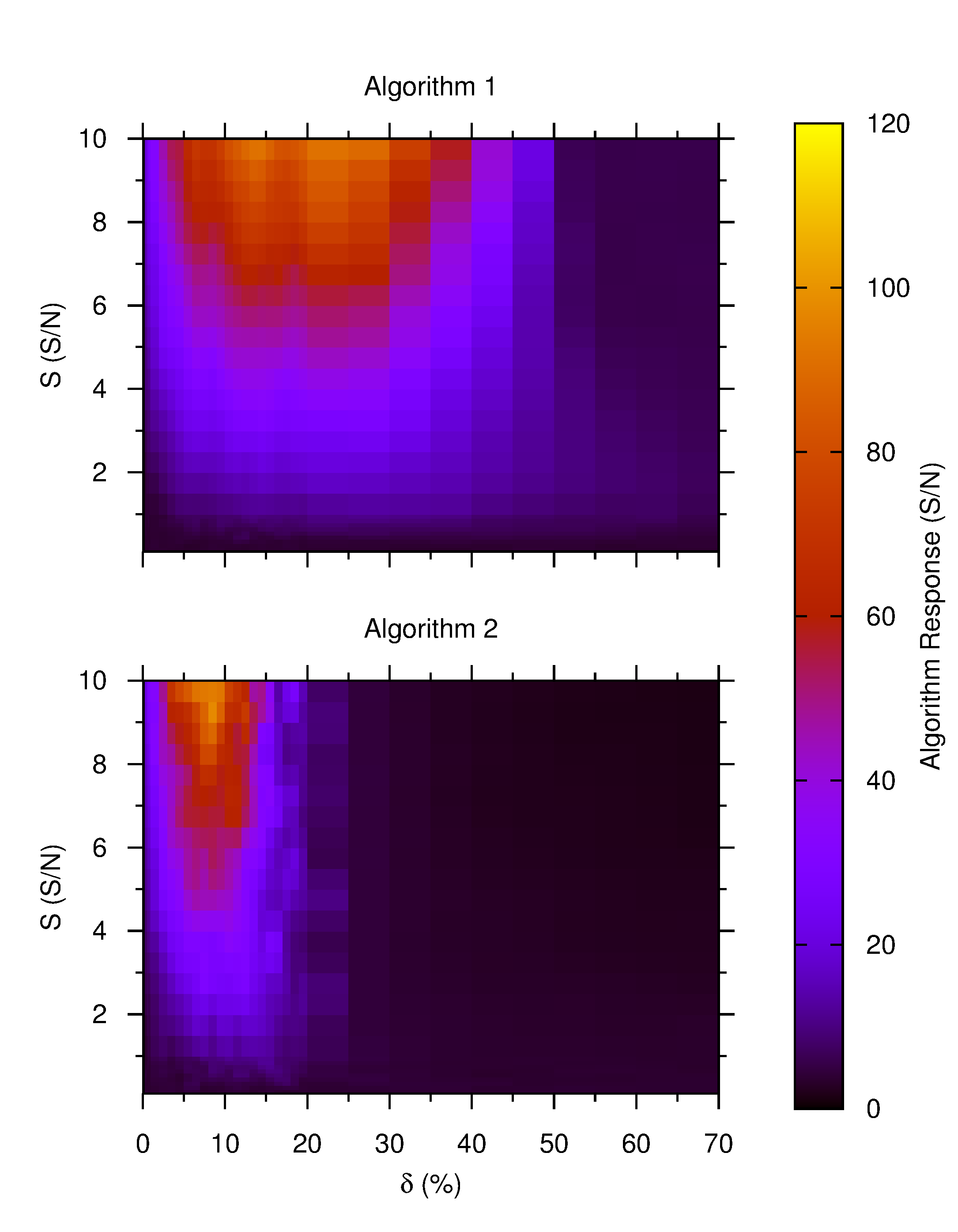}
\caption{Response patterns for Algorithms 1 \& 2 in relation to variations in pulse height $S$ (in units of the standard deviation of the profile white noise) and duty cycle $\delta$.\label{fig:DCH Figure}}
\end{figure}

Based upon Equation \ref{Single pulse FFA sensitivity - DCH}, for a constant period, the response of both algorithms should increase both with increasing duty cycle $\delta$ and with increasing pulse height $S$. As a reflection of the overall pulsar population \citep[with statistics derived using the ATNF Pulsar Catalog \texttt{psrcat}\footnote{http://www.atnf.csiro.au/research/pulsar/psrcat/},][]{mhth05}), duty cycle ranges of $0.1\%$ to $1.0\%$ (increments of $0.1\%$), $1.0\%$ to $20.0\%$ (increments of $1.0\%$) and $20.0\%$ to $70.0\%$ (increments of $5.0\%$) were selected for testing, comprising a total of 39 test values (although pulsars with smaller duty cycles than $0.1\%$ are known, $0.1\%$ was chosen as the lower limit based upon the number of bins used in the test profiles). The pulse height $S$, specified as multiples of the standard deviation of the white noise superimposed onto the profile, was tested in a range from $0.1$ to $1.0$ (increments of $0.1$) and from $1.0$ to $10.0$ (increments of $0.5$), comprising a total of 28 test values. These values were chosen as representative of weaker pulsar detections, which our work with the FFA hopes to improve. Altogether, the probed parameter space consists of 1092 test cases. The results of the simulation can be found in Figure \ref{fig:DCH Figure}.

From these results, several key features can immediately be identified, with the two algorithms displaying markedly different response patterns. Both algorithms display a similar region of maximum response, with Algorithm 2 reaching a slightly higher maximum than Algorithm 1, but this region is constrained differently in each case. In line with our earlier prediction, at constant pulse height the algorithm response increases with increasing duty cycle, but only up to a specific point. Algorithm 1 reaches a maximum at approximately $\delta=15 - 25\%$ and then gradually weakens towards higher duty cycles, whereas Algorithm 2 reaches a maximum much earlier at approximately $\delta=7-8\%$, before sharply falling off at $\delta=12 - 13\%$ and practically vanishing above $\delta=20\%$. 

In the case of Algorithm 1, this deviation can be explained as a consequence of the fact that Algorithm 1 ceases applying increasingly large matched-filters once their size exceeds $\delta=20\%$. Hence, the algorithm loses response as pulse sizes exceed this maximum matched-filter. Additionally, increasingly wide pulses will begin to render the MAD normalisation scheme less appropriate. A wider pulse will both raise the value of the median, resulting in an incorrect baseline correction with a significant fraction of the pulse sitting below the median, as well as a raising the approximated standard deviation of the profile, further decreasing Algorithm 1's responsiveness. However, the scale of this effect is sufficiently small to allow for stronger pulses to be detected up to duty cycles of $\delta\simeq50\%$.

In the case of Algorithm 2, in addition to using the same limited number of matched-filter sizes as used in Algorithm 1, it also applies a $20\%$ duty cycle exclusion window. This is likely responsible for the vanishing response that occurs at approximately that duty cycle value. Pulses wider than this exclusion window will begin to ``contaminate'' the assumed off-pulse statistics for the average intensity and standard deviation, introducing the same response restrictions as identified in Algorithm 1, but with the degradation occurring at lower values of duty cycle. However, counter-intuitively the loss of response begins at duty cycles as low as $\delta=12 - 13\%$, smaller than the $20\%$ duty cycle exclusion window would suggest. This is likely a result of the maximum value in the top hat profile being detected towards one edge, a result of the random variations introduced by the superimposed profile noise. This leads to the algorithm incorrectly centering the exclusion window, leaving some of the pulse outside of the window and thus contaminating the baseline statistics.

Finally, in the region where the value of $\delta$ allows for a strong detection, our second prediction regarding the algorithm response also holds. That is, as $\delta$ is held constant, the response of both algorithms increases with increasing pulse height $S$.

\subsubsection{Duty cycle vs. pulse energy}\label{subsec: DCE} 

\begin{figure}
 \includegraphics[width=\columnwidth]{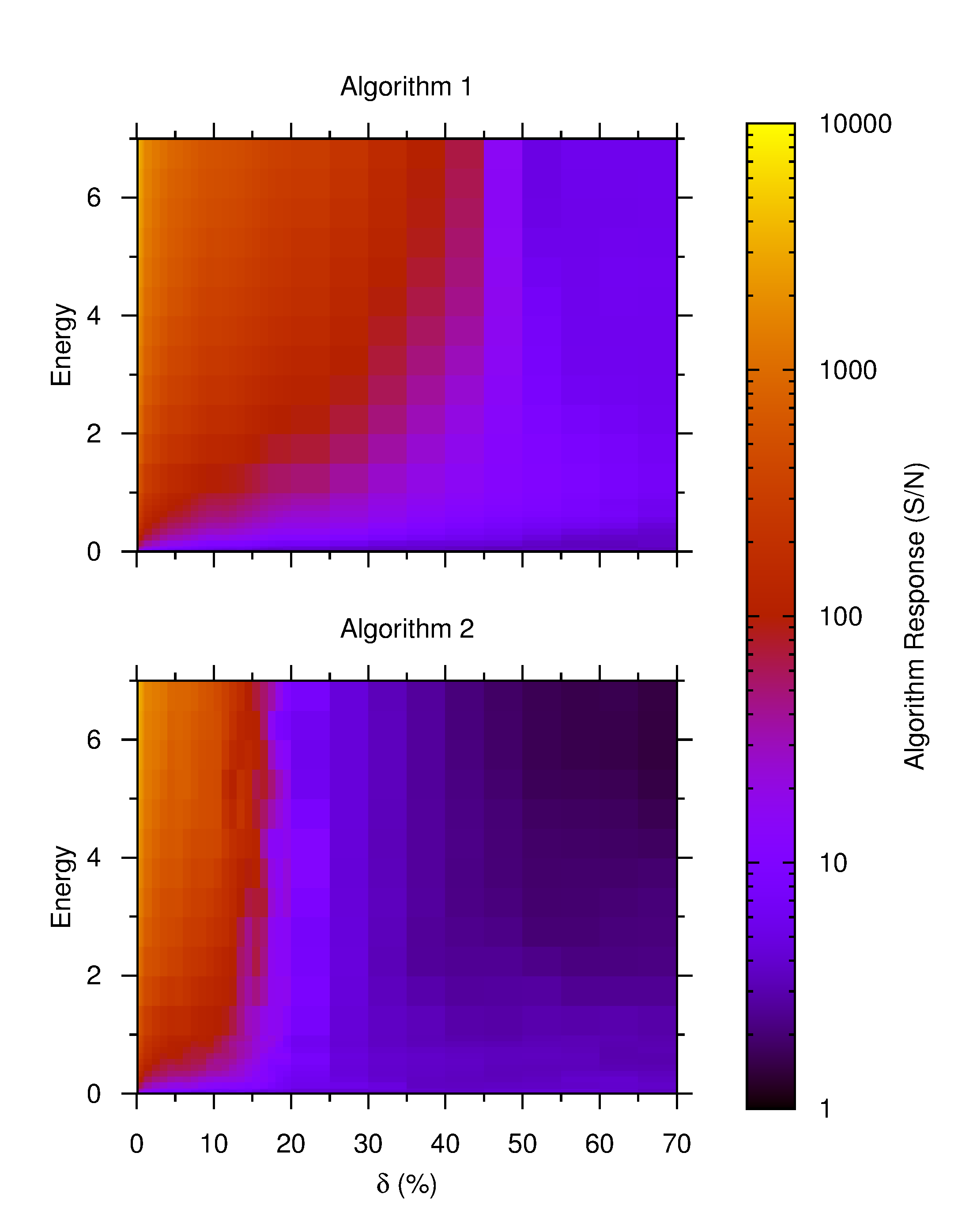}
\caption{Response patterns for Algorithms 1 \& 2 in relation to variations in duty cycle $\delta$ and energy. Note that the colour scale in this figure is logarithmic, due to the inversely proportional relationship between $\delta$ and algorithm response.}\label{fig:DCE Figure}
\end{figure}

As with duty cycle and pulse height, Equation \ref{Single pulse FFA sensitivity - DCE} allows us to predict the effects of manipulating the duty cycle and pulse energy, namely that the algorithm response should increase with increasing energy $E_p$ and should decrease as the duty cycle $\delta$ increases. By Equation \ref{E=wS}, the values of $E_p$ probed in Section \ref{subsec: DCH} range between $0.0001$ and $7.0$ (units arbitrary). In order to probe the same parameter space, test values were chosen between $1.0$ and $7.0$ (increments of $0.5$), $0.1$ and $1.0$ (increments of $0.1$) and $0.01$ and $0.1$ (increments of $0.01$), with additional tests at $0.006$, $0.001$, $0.0006$ and $0.0001$, comprising 35 test values. The same range of duty cycles as tested in Section \ref{subsec: DCH} was re-used, producing a parameter space consisting of 1365 test cases, each of which was repeated 30 times with the average taken as the final result. The results of the simulation can be found in 
Figure \ref{fig:DCE Figure}.

% energy test cases at the lower end may need to be slightly refined

From these results, similar trends as observed during the duty cycle / pulse height trials are immediately apparent. As before, Algorithm 1 and Algorithm 2 display similar yet different response patterns, Algorithm 1 displaying a broader response pattern while Algorithm 2 displays a sharper, narrower response. The same limitations in duty cycle response as noted in Section \ref{subsec: DCH} apply here, and will not be rediscussed. At smaller duty cycles unaffected by these limitations, it can be seen that, as predicted, both algorithm response patterns increase dramatically as the pulse energy increases. It would seem clear that both algorithms prefer tall, narrow pulses rather than broader, shallower pulses with the same energy. Additionally, Algorithm 2 once again produces a marginally stronger maximum response than Algorithm 1. This is consistent with the earlier difference in algorithm strengths displayed in Section \ref{subsec: DCH}.

\subsubsection{Pulse height vs. pulse energy}\label{subsec: EH}

\begin{figure}
  \includegraphics[width=\columnwidth]{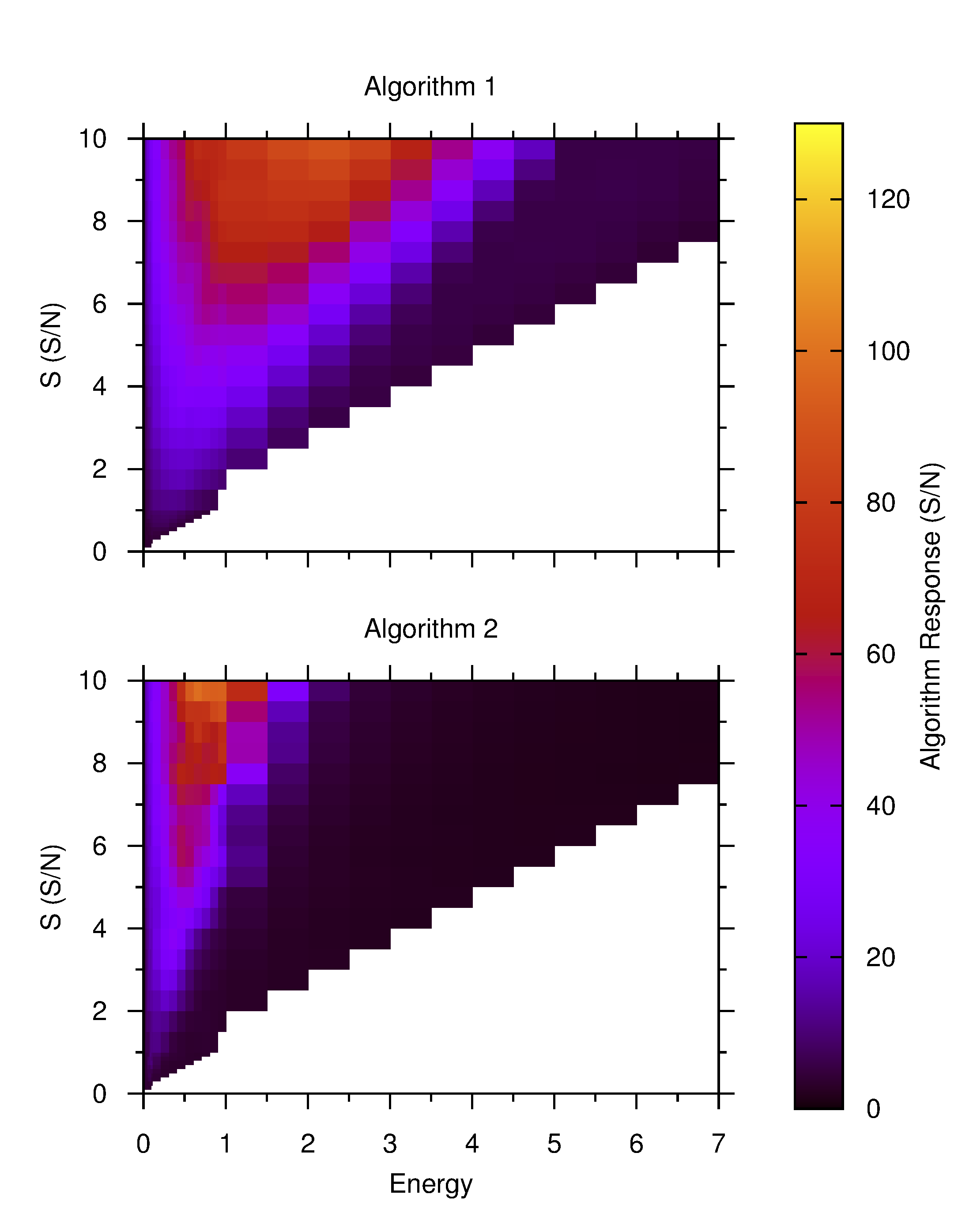}
\caption{Response patterns for Algorithms 1 \& 2 in relation to variations in pulse height and energy. The empty regions in the bottom right of each plot represent regions where the duty cycle $\delta\geq100\%$. This region was therefore excluded from testing.\label{fig: EH Figure}}
\end{figure}

Finally, using Equation \ref{Single pulse FFA sensitivity - EH} we can predict the behaviour of both algorithms as we vary pulse height and pulse energy in tandem. As both energy and height are increased, the response of both algorithms should also increase. The tested values for both pulse energy $E_p$ and height $S$ are those tested in Sections \ref{subsec: DCH} and \ref{subsec: DCE}, giving a total parameter space consisting of 980 test cases, each of which was repeated 30 times with the average taken as the final result. However, some combinations had to be excluded where the combination would result in a duty cycle of $\delta\geq100\%$. The results of the simulation can be found in Figure \ref{fig: EH Figure}.

The displayed response patterns and relative strengths between the two algorithms are in line with those from the previous simulations. Holding $E_p$ constant, the response of both algorithms does indeed increase with increasing pulse height, but while increasing $E_p$ at constant pulse height $S$ does initially cause an increase in response, this response falls off as $E_p$ is further increased. This is because increasing the pulse energy at constant height must necessarily result in a larger duty cycle $\delta$, re-introducing the same limitations as were previously encountered.

\subsection{Period}\label{subsec:Period}

\begin{figure*}
 	\includegraphics[width=\textwidth]{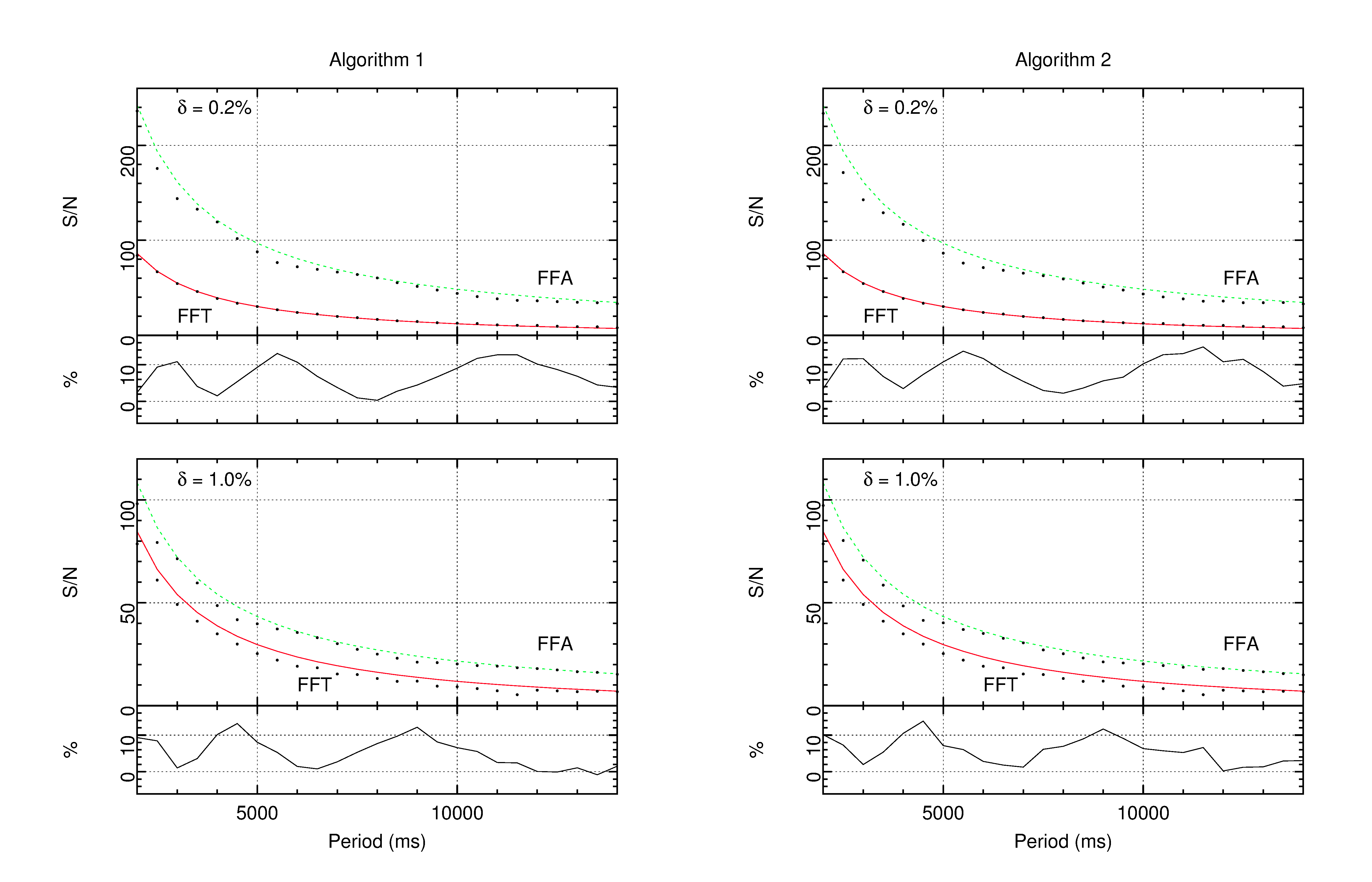}
    \caption{Simulation results demonstrating the performance of Algorithms 1 \& 2 against the ideal FFA response and the FFT at two different duty cycles. In each figure, the solid curves represent the ideal analytical response while the point represent the experimental results of the FFA and FFT respectively. The curve beneath each plot represents the percentage deviation of each algorithm from the analytical FFA response.}
    \label{fig:M6 and M7 period tests}
\end{figure*}

To test the performance of the algorithms as a function of period, we return to the simulation conducted in Section \ref{KondratievCorrectness}. Using the same test cases and procedure (including constant pulse energy between trials), we can produce similar FFA performance plots for both Algorithm 1 and Algorithm 2. The results of this simulation can be found in Figure \ref{fig:M6 and M7 period tests}.

From these results, it can be seen that both algorithms obey the same trends predicted by Equation \ref{FFA_sensitivity}, i.e. with constant energy, duty cycle and observation length, the response of each algorithm should increase inversely proportional to the period $P$. Also, although the performance of both algorithms degrades slightly from the ideal analytical response, they still maintain a higher S/N result across both example duty cycles ($\delta=0.2\%$ and $1.0\%$) and all periods than the 16 harmonic sum FFT. This degradation from the analytic FFA curve is limited to a maximum of around $15-16\%$, as indicated by the fractional deviation plots located beneath each S/N plot. As can be seen in each of the examples, these deviations follow a rough cycle whose period increases by a factor of 2 with each oscillation, with the shape of the curve apparently common to both algorithms for each duty cycle. This oscillation is linked to the way in which the matched-filter operates in both algorithms, increasing in size by factors of 2 such that the matched-filter whose width is closest to that of the signal in the profile should return the highest response. Minima in the deviation plots correlate with pulse widths relatively close to the size of one of these trial matched-filters, while maxima correlate with pulse widths caught almost precisely between two matched-filter sizes, neither able to truly capture the width of the pulse.

\subsection{Pulse Shape}
For simplicity, all tests thus far have been conducted using top-hat pulse profiles. However, although it is a useful approximation, this pulse shape is unrealistic when considering real pulsars. The morphology of pulsar profiles can range from simple single Gaussian-like curves to much more complicated shapes, and these varying profiles may influence the response of the algorithms. The taxonomy scheme proposed by \cite{bac76} and refined by \cite{ran83} classifies pulse profiles into several broad categories according to the apparent shape of profile components and their associations with models of pulsar emission. However, as we are interested purely in the effects of variations in the pulse shape as well as the number of components, we can employ a simpler taxonomy scheme in our consideration of which profile types to investigate. These tests must also allow for shape variations caused by the presence of scattering \citep[e.g.][]{bcc+04} and interpulses \citep[e.g.][]{lk05}, which can further modify the shape of the observed profile and may affect their detectability using our chosen algorithms. Given the considerable size of the parameter space that could be explored in a full investigation of pulse shape variance, the tests chosen here are not intended to be exhaustive, but will instead serve as case examples reflective of pulsars likely to be encountered during an FFA search. The results of these tests are detailed in the following subsections.

\subsubsection{Single-component Gaussian profile}\label{subsubsec: single Gaussian} 

\begin{figure}
	\includegraphics[width=\columnwidth]{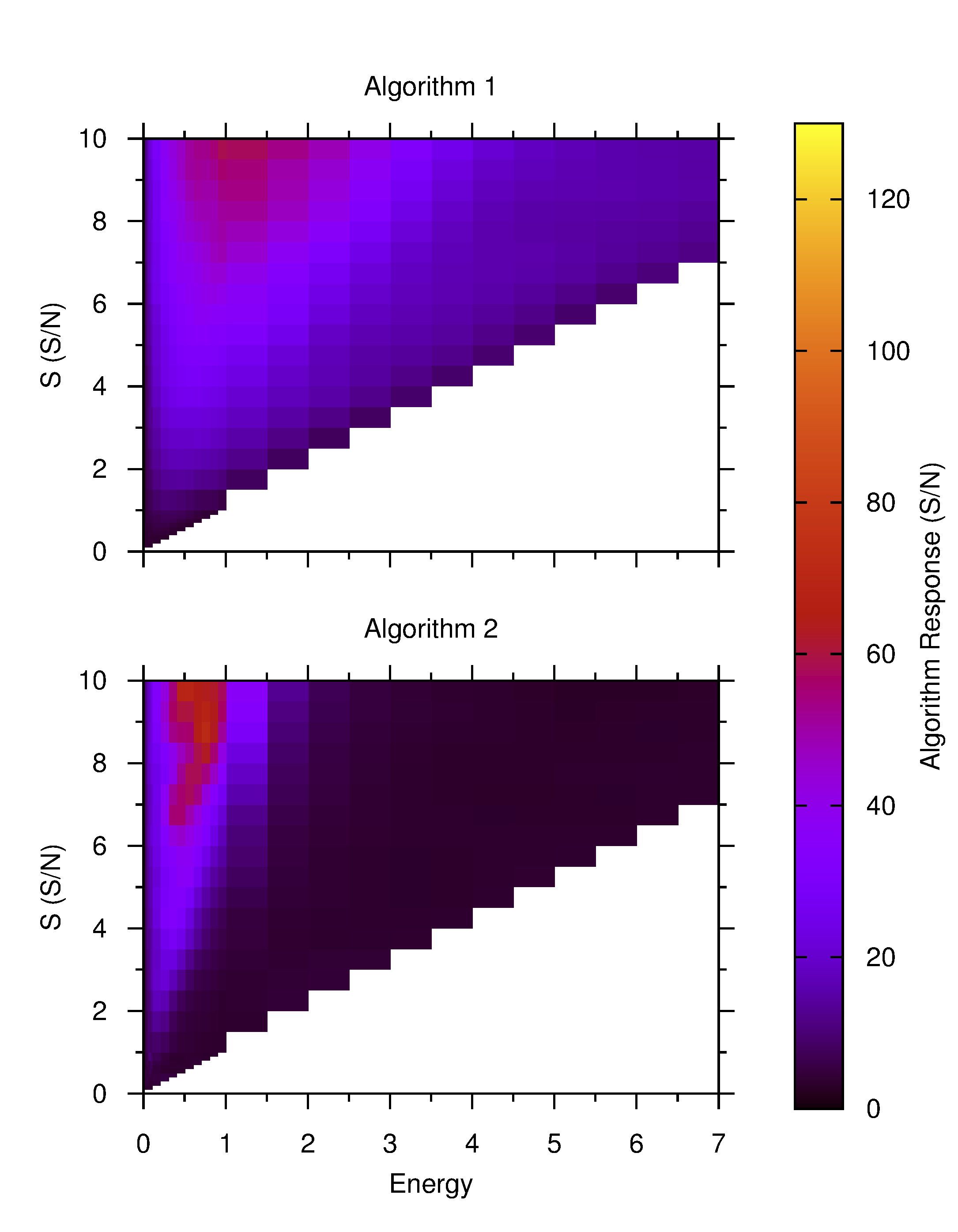}
  \caption{Identical to the simulation results displayed in Figure \protect\ref{fig: EH Figure}, but for a single-component Gaussian profile instead of the original top-hat profile.}
 \label{fig: EH Gaussian}
\end{figure}

The most obvious first step in implementing a more realistic pulse shape is to change from a top-hat pulse to a Gaussian, as this is an approximation appropriate to many known pulsars. This change in shape modifies the earlier expressions dervied for the pulse width and energy. For a Gaussian profile with maximum amplitude $S$ and variance $c > 0$, an appropriate substitute for the duty cycle $\delta$ becomes the FWHM, defined by\begin{equation}\label{eqn: FWHM}
                                                                                                                                                                                                                                                                                                                                                                                                                                                                   \delta = \text{FWHM} = 2\sqrt{2\ln{2}}c.
                                                                                                                                                                                                                                                                                                                                                                                                                                                                  \end{equation} The energy of the pulse $E_g$ similarly becomes \begin{equation}\label{eqn: gauss energy}
                                                                                                                                                                                                                                                                                                                                                                                                                                                                                                                            E_g = Sc\sqrt{2\pi} = \frac{S\delta\sqrt{\pi}}{2\sqrt{\ln{2}}}.
                                                                                                                                                                                                                                                                                                                                                                                                                                                                                                                           \end{equation} 
                                                                                                                                                                                                                                                                                                                                                                                                                                                                                                                           To compare the response of both algorithms to Gaussian pulse shapes, we repeated the simulation conducted in Section \ref{subsec: EH}, using the same energy and height combinations with 30 repetitions per test, with the average of each test taken as the final result. Energy and maximum pulse height were chosen as the defining parameters of the test space as they are both easier to define in the case of a Gaussian pulse than the duty cycle, a problem which will be only be exacerbated with the more complex pulse shapes encountered in later tests. The results of this simulation can be found in Figure \ref{fig: EH Gaussian}. It can be seen that in comparison to the top-hat simulation in Figure \ref{fig: EH Figure} the overall response of both algorithms is similar in terms of its distribution across the energy-height plane, likely for the same reasons as discussed in Section \ref{subsec: EH}, but in both cases each algorithm gives an overall lower magnitude of response (note that both figures employ an identical colour scale for comparison).

\subsubsection{Single-component Gaussian profile in the presence of scattering}\label{subsubsec: single Gaussian scattered}

Now that a baseline response to a single-component Gaussian profile has been established, the responsiveness of both algorithms to the presence of an increasingly scattered Gaussian pulse can also be investigated. The effect of scattering can be simulated by convolving the original pulse shape with a one-sided exponential decay function with a scattering time scale of $\tau_s$ \citep{lk05}. As the profiles generated in these series of tests have been independent of any fixed pulsar period, $\tau_s$ is in units of profile bins. Since the effects of scattering often smear out and obscure any more complex pulse shape features \citep{bcc+04}, it is sufficient to conduct this test with a single-component Gaussian profile.

\begin{figure}
 \includegraphics[width=\columnwidth]{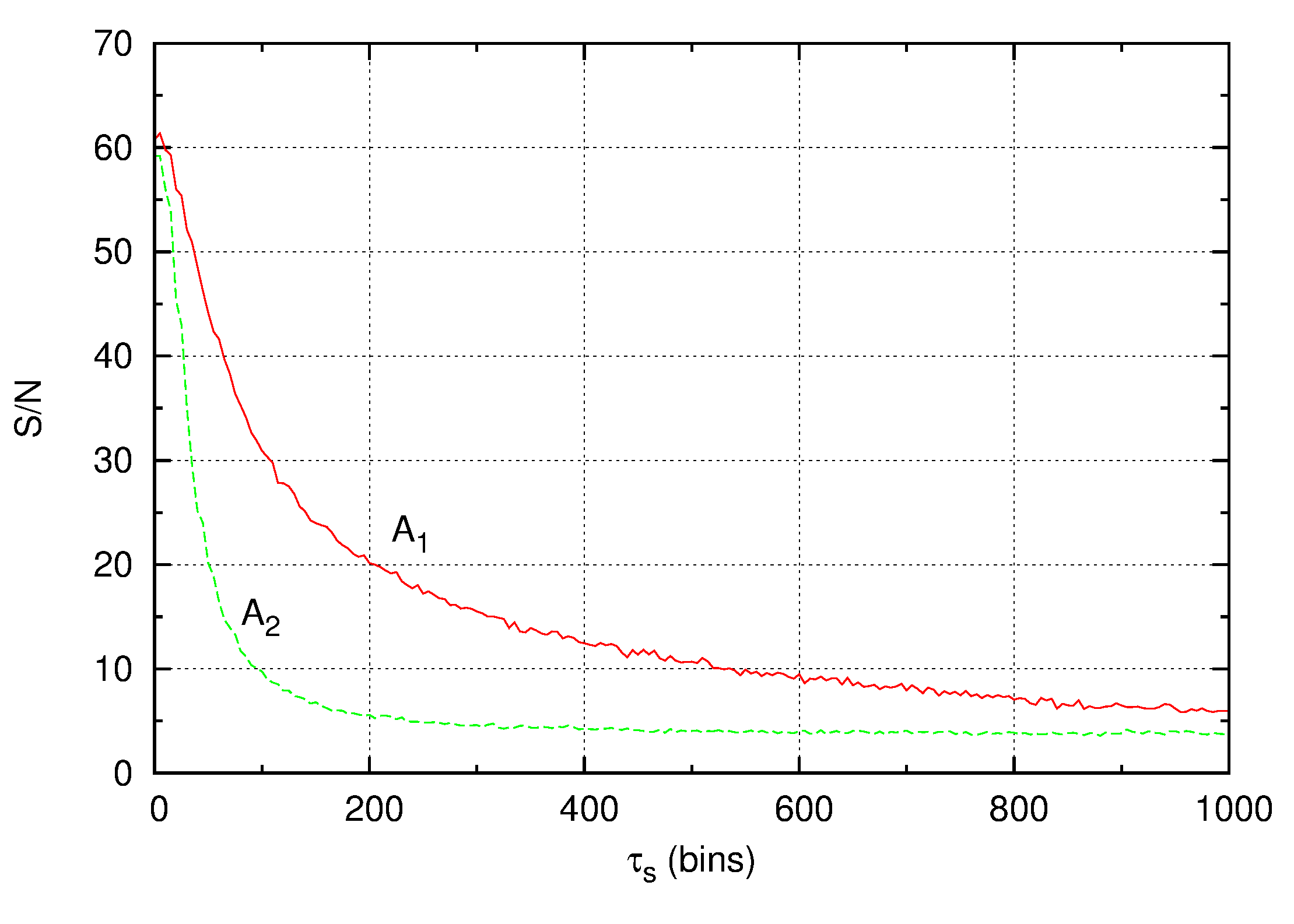}
  \caption{Algorithm response to an increasingly scattered single Gaussian pulse. Original pulse seeded with height 10 times the profile noise and $\delta\simeq9.39$, producing a pulse energy $E_g=1$.}
 \label{fig: gauss scattering}
\end{figure}

To investigate the response of both algorithms to an increasingly scattered profile, a single test case was selected from the single-component Gaussian simulation which produced a high relative response in both algorithms, i.e., an unscattered pulse height of $S = 10$ times the profile noise level and energy $E_g = 1$. By Equation \ref{eqn: gauss energy}, this produces a duty cycle of $\delta\simeq 9.39\%$. Test profiles were generated using values of $\tau_s$ ranging from $0$ to $1000$ profile bins in increments of $5$. As the number of bins in the profile is 1024, the maximum value of $\tau_s$ represents a pulse that has been almost completely scattered across the width of the profile. This resulted in 201 test cases, each of which was repeated 30 times. The average response of these 30 repetitions was taken as the final result for each test case. In each test case, it was ensured that the energy of the scattered pulse remained consistent with the energy of the unscattered pulse. The results of the simulation can be found in Figure \ref{fig: gauss scattering}.

It can be seen that both algorithms display markedly different responses to the presence of increased scattering. At very low values of $\tau_s$ both algorithms essentially provide the same response, but as the value of $\tau_s$ increases, Algorithm 2 exhibits a much more dramatic fall in response as compared to the more gentle decline in response of Algorithm 1. By $\tau_s\simeq200$, Algorithm 2 has almost plateaued at a response of $A_2\simeq4$, while Algorithm 1 only begins to approach this response at the highest values of $\tau_s$ tested. This difference in behaviour between the algorithms is likely another reflection of the same difference in response exhibited to varying duty cycles as seen in prior tests. As the Gaussian pulse is increasingly scattered, it both widens asymmetrically and its maximum amplitude decreases in order to keep the pulse energy constant. Given the $20\%$ exclusion window imposed by Algorithm 2, it exhibits a much sharper decline in response as the pulse scatters outside of this window.

\subsubsection{Two-component Gaussian profiles}

\begin{figure}
  \includegraphics[width=\columnwidth]{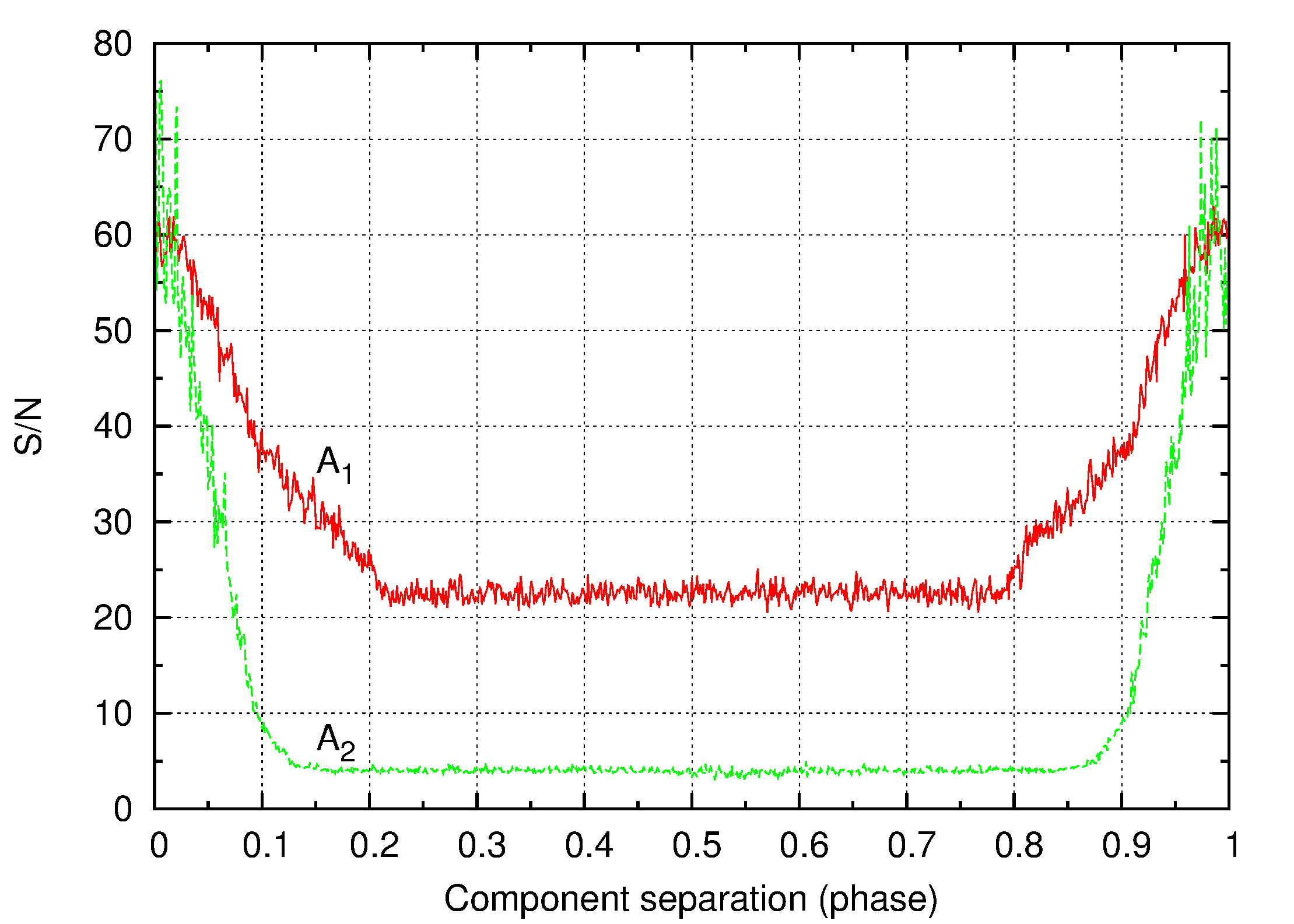}
  \caption{The response of Algorithms 1 \& 2 to a two-component Gaussian profile, with both components being identical. The central position of the second pulse was shifted across the entire width of the profile.}
 \label{fig: doublegaussequal}
\end{figure}

To investigate the response of both algorithms to the presence of more complicated profiles, two further simulations were performed involving two-component Gaussian profiles. In the first simulation, profiles containing two Gaussian pulses of identical strength were generated, with the separation between these pulses varied. Such a simulation encompasses the response of the algorithms both to more complicated, two-component pulse profiles (when the pulse peaks are located close together) and to harmonic detections of a pulsar folded at the wrong period (when the pulse peaks are located further apart). To produce the simulation, profiles were created containing two Gaussian pulses. Each pulse was seeded with a height of $S = 10$ times the profile noise level and energy $E_g = 0.5$. By Equation \ref{eqn: gauss energy}, each pulse would again have a duty cycle of $\delta\simeq 9.39\%$, and when added together would produce the same unscattered pulse as used in the scattering trials. One pulse was held stationary while the other was moved such that its peak was trialled at each bin position in the 1024 bin profile. Each bin position test was repeated 30 times, with the average taken as the final result for each bin position test. The results of this first simulation can be found in Figure \ref{fig: doublegaussequal}.

\begin{figure}
  \includegraphics[width=\columnwidth]{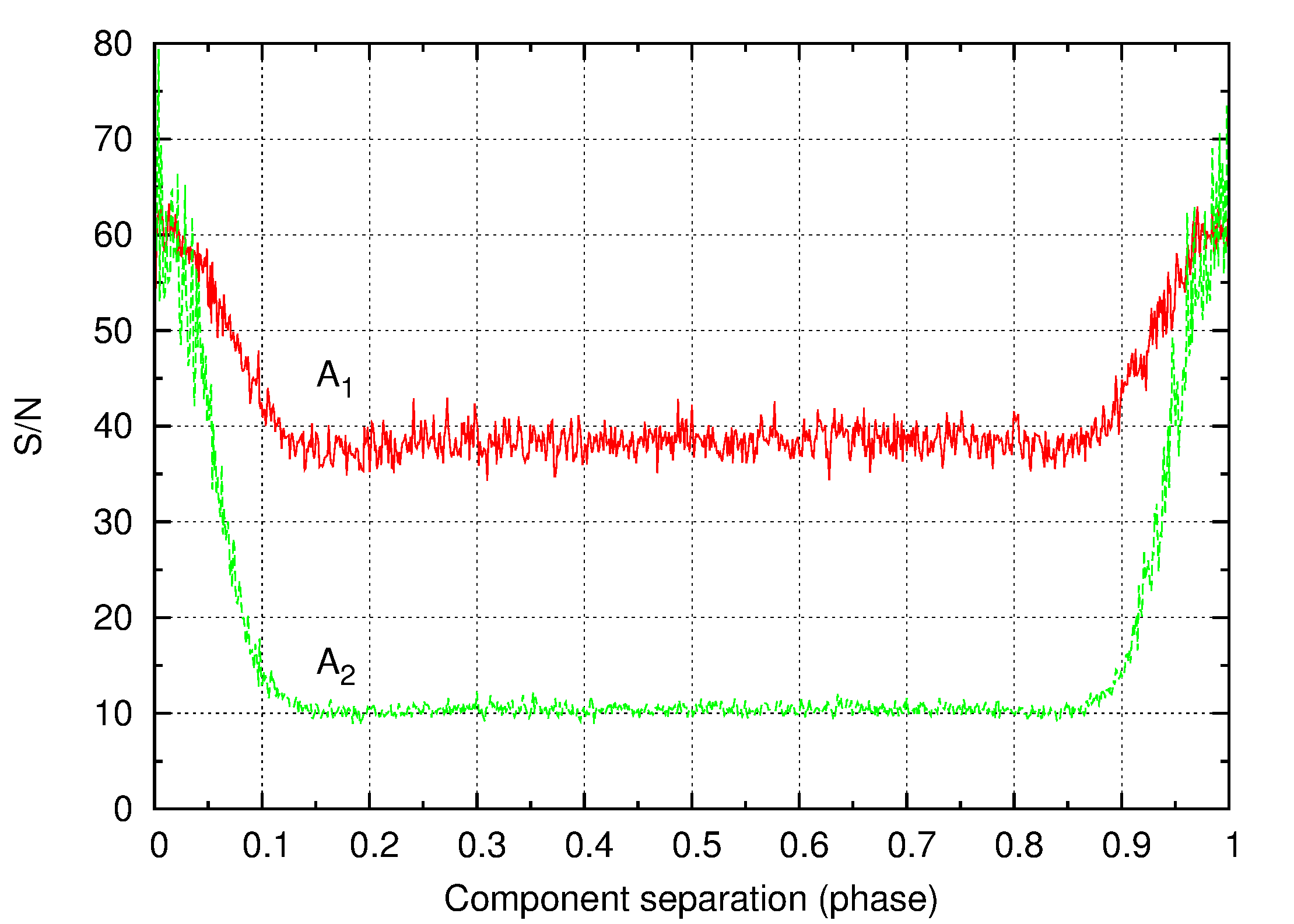}
  \caption{The response of Algorithms 1 \& 2 to a two-component Gaussian profile, with both components being of unequal height and energy. The central position of the second, weaker pulse was shifted across the entire width of the profile.}
 \label{fig: doublegaussunequal}
\end{figure}

In the second simulation, profiles containing two Gaussian pulses of unequal strength were generated, with the separation between these pulses once again varied. Such a simulation again encompasses the response of the algorithms both to more complicated, two-component pulse profiles (when the pulse peaks are located close together) as well as to the presence of interpulses (when the pulse peaks are located further apart). The parameters of the test were identical to that of the previous simulation, with the exception that the stationary pulse was seeded with $S = 7.5$ and $E_g=0.75$, while the moving pulse was seeded with each of these parameters divided by 3, such that $S = 2.5$ and $E_g=0.25$. This would again result in two pulses with $\delta\simeq 9.39\%$ and that when added together would reproduce the unscattered pulse trialled as used in the scattering trials. The results of this simulation can be found in Figure \ref{fig: doublegaussunequal}.

Familiar trends can be observed in these results. Once again, as the two peaks are gradually separated, creating a wider profile with multiple components, the response of both algorithms decreases in both test cases, with Algorithm 2 decreasing and plateauing before Algorithm 1 for the reasons already outlined in earlier sections. The more interesting result can be seen in the center of the both Figures, where the two pulse components have moved far enough apart to be treated as separate features. It is clear that in this region both algorithms make no distinction about the location of the second pulse, maintaining a flat response throughout, but both algorithms plateau at different levels. Algorithm 1 maintains a stronger response in this region than Algorithm 2. This is likely due to the fact that the MAD normalisation technique is less affected by outliers (in this case the additional pulse component) than the standard deviation technique used in the evaluation of Algorithm 2. This contamination is reduced in Figure \ref{fig: doublegaussunequal}, as the second, moving pulse is relatively weaker in this case, allowing for the flat response of both algorithms to rise.

\section{Trials on real observational data}\label{sec:realtrial}

All testing thus far has focused on evaluating the behaviour of the FFA and its associated algorithms in the white noise regime. In order to quantify the response of the FFA to the presence of red noise and other RFI and to compare its response to that of the FFT, a trial was conducted using real observational data taken from the High Time Resolution Universe South Low Latitude survey \citep[HTRU-S LowLat,][]{kjvs10}, conducted with the Parkes 64-m radio telescope. 20 telescope beams containing a selection of the longest period known pulsars present in the survey area were selected. These pulsars cover a period range from approximately $1.2$ to $6.4 s$, and represent a significant range of apparent flux densities and pulse shapes. For example, J1307-6318 displays a striking two-component profile, while the magnetar J1622-4950 displays a very wide profile (with a duty cycle as high as $\delta \simeq 50-60\%$) which is known to vary with time \citep{lbb+10}. Table \ref{tab:HTRU Pulsar Table} lists the details of these pulsars as obtained through \texttt{psrcat}.

\begin{table*} 
\caption{Parameters of the test pulsars selected from the HTRU-S LowLat survey as provided by \texttt{psrcat}, listed from longest to shortest barycentric period, along with the derived red noise properties of their respective observations. The Pointing/Beam combination identifies which survey observation and beam number that the pulsar was recorded in. The expected S/N is calculated based on the radiometer equation, and takes into account the pulsar's non-central position in the beam. Duty cycle $\delta$ was derived using the FWHM (where available) as $\delta = \text{FWHM}/P$. The red noise properties of each observation were determined through the use of a power law fit, with the amplitude and index of this power law (as well as the frequency below which the power law fit was appropriate) listed.}
\label{tab:HTRU Pulsar Table}
  \begin{center}
   \begin{tabular}{cc|cccc|ccc}
  \hline
   & & \multicolumn{4}{c}{Pulsar properties} & \multicolumn{3}{c}{Red noise properties} \\
  PSR name & Pointing/Beam & $P \text{ (ms)}$ &  $DM \text{ (cm}^\text{-3}\text{pc)}$ & $\delta \text{ (\%)}$ & S/N & Amplitude & Index & Cutoff Freq. (Hz)\\ \hline
  J1736-2843 & 2011-07-03-13:48:57/05 & 6445.036 & 331 & 2.25 & 25.1 & 35.86 & -1.86 & 1.07 \\
  J1840-0840 & 2013-01-01-02:30:42/10 & 5309.377 & 272 & 3.39 & 151.2 & 22.92 & -2.00 & 0.90 \\
  J1307-6318 & 2011-12-29-14:57:57/08 & 4962.427 & 374 & 10.18 & 20.7 & 12.91 & -2.12 & 0.67 \\
  J1622-4950 & 2010-12-31-21:03:27/08 & 4326.100 & 820 & - & 52.2 & 28.49 & -2.03 & 1.20 \\
  J1814-1744 & 2013-01-01-00:04:30/08 & 3975.905 & 792 & 2.31 & 28.3 & - & - & - \\
  J1718-3718 & 2011-06-27-14:48:40/13 & 3378.574 & 371.1 & 3.85 & 19.1 & 27.54 & -1.77 & 1.93 \\
  J1314-6101 & 2011-06-30-10:40:03/03 & 2948.390 & 309 & 2 & 69.6 & 28.39 & -2.25 & 1.50 \\
  J1803-1857 & 2012-02-16-01:37:19/02 & 2864.338 & 392 & 0.89 & 98.3 & 98.30 & -1.93 & 3.79 \\
  J1831-1223 & 2011-12-06-05:56:02/12 & 2857.941 & 342 & 3.43 & 63.9 & 36.09 & -1.78 & 2.63 \\
  J1759-1956 & 2012-02-16-01:37:19/12 & 2843.389 & 236.4 & 1.33 & 59.2 & 94.92 & -1.96 & 3.61 \\
  J1747-2802 & 2011-05-17-15:33:25/07 & 2780.079 & 835 & 0.98 & 14.7 & 5.95 & -1.76 & 0.99 \\
  J1444-5941 & 2011-06-26-05:10:45/01 & 2760.228 & 177.1 & 1.7 & 17.4 & 37.51 & -1.88 & 2.50 \\
  J1811-1049 & 2012-04-01-18:52:53/05 & 2623.859 & 253.3 & - & 5.3 & 9.97 & -2.01 & 1.19 \\
  J1801-1855 & 2012-02-16-01:37:19/07 & 2550.498 & 484 & 1.64 & 86.1 & 76.22 & -1.90 & 3.85 \\
  J1822-0848 & 2012-04-02-18:07:28/13 & 2504.518 & 186.3 & 1.24 & 1.74 & 14.88 & -1.99 & 1.55 \\
  J1324-6302 & 2011-12-08-02:26:00/13 & 2483.804 & 497 & 0.72 & 43.0 & 6.21 & -1.91 & 1.05 \\
  B1658-37 & 2012-04-02-14:27:04/07 & 2454.609 & 303.4 & 1.75 & 393.0 & 9.06 & -1.98 & 1.24 \\
  B1740-31 & 2011-10-13-10:01:19/13 & 2414.576 & 193.05 & 1.86 & 207.0 & - & - & - \\
  J1817-1938 & 2013-04-08-18:54:20/10 & 2046.838 & 519.6 & - & 15.6 & 1.67 & -2.03 & 0.63 \\
  J1838-1046 & 2011-07-14-15:08:44/07 & 1218.354 & 208 & 1.36 & 15.5 & 9.98 & -2.01 & 2.58 \\ \hline 
  \end{tabular}
    
  \end{center}
\end{table*} 

Processing was undertaken once again using software from the \texttt{SIGPROC} software package. Each filterbank observation file was RFI-cleaned using a process of channel and timezapping as adapted from \cite{ncb15}. Each filterbank was then dedispersed using the value of DM appropriate for its pulsar to produce a timeseries. The red noise present in each time series was characterised through the use of a power law fit, with the index of this power law and the frequency below which the power law fit was appropriate being listed in Table \ref{tab:HTRU Pulsar Table}. Each timeseries was then processed through an FFT and FFA analysis in the following manner:
\begin{itemize}
 \item \textbf{FFT:} Each timeseries was processed through the \texttt{SIGPROC} program \texttt{seek}, with the use of the default AGN spectral-whitening dereddening scheme and the summation of 16 harmonics. A spectral bin-zapping mask, or ``birdie'' mask, was also applied. This mask takes advantage of the 13 simultaneous beams recorded during each HTRU-S LowLat observation, and was generated through the use of the multibeam RFI-excision technique described in \cite{ncb15}. The resulting candidate file was then evaluated by eye to find the strongest detection which was harmonically related to the relevant pulsar.
 \item \textbf{FFA:} Each timeseries was processed through \texttt{ffancy} using a search from $1$ to $20\text{ s}$. An initial downsampling factor of $2^6 = 64$ was applied to each time series before the search began, so as to reduce the required processing time. \texttt{ffancy}'s dynamic dereddening scheme was activated for each timeseries, using the default values described in Section \ref{subsubsec:ffancy software}. This process was repeated with both Algorithm 1 and Algorithm 2. Finally, the resulting periodograms for each pulsar and algorithm were processed through the candidate selection program \texttt{ffa2best}. No candidate was accepted with an S/N of less than 9, and both candidate grouping and harmonic matching were used so as to lower the number of overall candidates. Harmonic matching was conducted using the complete set of primes up to and including 31, with a matching tolerance of $0.001\%$ for Algorithm 1 and $0.01\%$ for Algorithm 2. These values were chosen as during testing they produced an 
optimal balance between correctly grouping related candidates and incorrectly grouping unrelated candidates.
\end{itemize}
As a control test on each pulsar's detectability, each clean time series was also folded using an optimal ephemeris as obtained using \texttt{psrcat} and visually inspected. The optimally folded profile of each pulsar at its fundamental period as produced using Algorithm 1 of the FFA was also produced for comparison.

The best detections made by each technique are given in Table \ref{tab:Results Table}. It should be noted that the periods detected by both the FFT and the FFA do not align precisely with the recorded \texttt{psrcat} values in Table \ref{tab:HTRU Pulsar Table}. This is partly due both to the resolution of each searching technique, as well as the change in epoch between the catalog period and each HTRU observation. Also, in many instances, these detections were made at harmonics of the true pulsar period. In the case of the FFT, five detections were made at various harmonics, and these occurrences are indicated in Table \ref{tab:Results Table}. In the case of the FFA, Algorithm 1 produced three detections at half of the true pulsar period which were stronger than the fundamental detection, while Algorithm 2 produced six stronger harmonic detections, one of which was at the quarter period harmonic while the others were at the half period harmonic. These detection S/Ns are compared against the fundamental S/Ns in Table \ref{tab:Fundamental Table}.

\begin{table*}
\caption{Results from the analysis of all 20 pulsars by the FFT and FFA using both Algorithms 1 \& 2. For reference, the \texttt{psrcat} pulsar periods are also included. $^\dag$FFT harmonic detections. Due to the nature of the operation of \texttt{seek}, the fundamental S/Ns are unavailable. $^\diamondsuit$FFA harmonic detections. For a comparison of these detections to the fundamental detections made by the FFA, refer to Table \protect\ref{tab:Fundamental Table}. All listed periods are topocentric.}
  \begin{center}
   \begin{tabular}{cccccccc}
  \hline
   & & \multicolumn{2}{c}{FFT} & \multicolumn{2}{c}{FFA (Algorithm 1)} & \multicolumn{2}{c}{FFA (Algorithm 2)}\\
  PSR name & $P \text{ (ms)}$ & $P \text{ (ms)}$ & S/N & $P \text{ (ms)}$ & S/N & $P \text{ (ms)}$ & S/N\\ \hline
  J1736-2843 & 6445.036 & 6445.421864 & 28.0 & 6445.254194 & 32.8 & 1611.298383$^\diamondsuit$ & 26.7 \\
  J1840-0840 & 5309.377 & 5309.085587 & 76.7 & 5309.441001 & 108.2 & 5309.200766 & 61.1 \\
  J1307-6318 & 4962.427 & 4962.142906 & 15.4 & 4962.334029 & 19.5 & 4962.253951 & 12.2 \\
  J1622-4950 & 4326.100 & 4326.605589 & 21.1 & 4326.513110 & 15.9 & - & - \\
  J1814-1744 & 3975.905 & 3976.188776 & 30.0 & 1988.062367$^\diamondsuit$ & 32.6 & 1988.058366$^\diamondsuit$ & 14.8 \\
  J1718-3718 & 3378.574 & 3378.580197 & 8.2 & 3378.917862 & 15.6 & - & - \\
  J1314-6101 & 2948.390 & 2948.479585 & 23.0 & 1474.271930$^\diamondsuit$ & 39.7 & 2948.527711 & 26.8 \\
  J1803-1857 & 2864.338 & 954.705686$^\dag$ & 31.1 & 2864.106490 & 52.1 & 2864.102488 & 35.6 \\
  J1831-1223 & 2857.941 & 2858.101443 & 51.6 & 2858.039527 & 50.5 & 1429.008879$^\diamondsuit$ & 31.2 \\
  J1759-1956 & 2843.389 & 1421.593545$^\dag$ & 17.6 & 2843.160262 & 30.4 & 2843.156260 & 20.2 \\
  J1747-2802 & 2780.079 & 2779.942215 & 14.6 & 2779.949397 & 14.7 & 2779.937391 & 11.3 \\
  J1444-5941 & 2760.228 & 2760.400348 & 9.3 & 2760.303805 & 13.9 & 2760.335820 & 11.3 \\
  J1811-1049 & 2623.859 & 2623.606783 & 11.1 & 2623.633071 & 17.7 & 2623.601055 & 15.7 \\
  J1801-1855 & 2550.498 & 1275.139081$^\dag$ & 21.6 & 2550.285256 & 34.3 & 2550.285256 & 24.9 \\
  J1822-0848 & 2504.518 & 1252.114308$^\dag$ & 9.0 & 2504.226767 & 14.4 & 2504.218763 & 11.4 \\
  J1324-6302 & 2483.804 & 620.930604$^\dag$ & 7.0 & 1241.868190$^\diamondsuit$ & 20.4 & 1241.861189$^\diamondsuit$ & 14.7 \\
  B1658-37 & 2454.609 & 2454.376590 & 116.6 & 2454.390433 & 304.8 & 2454.466470 & 178.2 \\
  B1740-31 & 2414.576 & 2414.874382 & 127.7 & 2414.851127 & 165.8 & 1207.415779$^\diamondsuit$ & 99.5 \\
  J1817-1938 & 2046.838 & 2046.638718 & 22.7 & 2046.615662 & 21.6 & 1023.330918$^\diamondsuit$ & 19.8 \\
  J1838-1046 & 1218.354 & 1218.392631 & 22.6 & 1218.376455 & 25.2 & 1218.377455 & 22.3 \\ \hline 
  \end{tabular}
    
  \end{center}
\label{tab:Results Table}
\end{table*}

\begin{table*}
\caption{Instances where the FFA detected a harmonic as the strongest detection using either Algorithm 1 or 2, as compared to the strength of the fundamental detection.}
  \begin{center}
   \begin{tabular}{ccccccccc}
  \hline
   & \multicolumn{4}{c}{FFA (Algorithm 1)} & \multicolumn{4}{c}{FFA (Algorithm 2)}\\
   & \multicolumn{2}{c}{Strongest Detection} & \multicolumn{2}{c}{Fundamental Detection} & \multicolumn{2}{c}{Strongest Detection} & \multicolumn{2}{c}{Fundamental Detection}\\
  PSR name & $P \text{ (ms)}$ & S/N & $P \text{ (ms)}$ & S/N & $P \text{ (ms)}$ & S/N\\ \hline
  J1736-2843 & - & - & - & - & 1611.298383 & 26.7 & 6445.222162 & 25.8 \\
  J1814-1744 & 1988.062367 & 32.6 & 3976.077444 & 23.0 & 1988.058366 & 14.8 & 3976.145477 & 8.7 \\
  J1314-6101 & 1474.271930 & 39.7 & 2948.527711 & 35.5 & - & - & - & - \\
  J1831-1223 & - & - & - & - & 1429.008879 & 31.2 & 2858.039527 & 25.8\\
  J1324-6302 & 1241.868190 & 20.4 & 2483. 732760 & 18.6 & 1241.861189 & 14.7 & 2483.728758 & 13.0 \\
  B1740-31 & - & - & - & - & 1207.415779 & 99.5 & 2414.775089 & 46.5\\
  J1817-1938 & - & - & - & - & 1023.330918 & 19.8 & 2046.621663 & 16.3\\ \hline 
  \end{tabular}
    
  \end{center}
\label{tab:Fundamental Table}
\end{table*}

\subsection{Analysis}\label{subsec:Analysis}

Both the FFT and Algorithm 1 of the FFA were able to successfully recover all 20 of the test pulsars. However, Algorithm 2 was only able to recover 18 of the 20 pulsars, with J1622-4950 and J1718-3718 remaining undetected. A likely explanation for the non-detection of J1622-4950 lies in the pulsar's unusually high duty cycle ($\delta \simeq 50-60\%$), because as demonstrated in Section \ref{subsec: DHE}, Algorithm 2 is largely insensitive to wide pulse profiles. Following a visual inspection of the folded profile produced by the FFA at the fundamental period, the non-detection of pulsar J1718-3718 appears to be due to a combination of an inherently low S/N combined with noise contamination of the baseline, which persisted despite the application of \texttt{ffancy}'s red noise removal system and other RFI mitigation techniques. The reduction in sensitivity caused by the contaminated baseline was again likely a result of Algorithm 2's insensitivity to wide pulsar profiles, or in this instance, wide baseline variations.

\begin{figure}
 \includegraphics[width=\columnwidth]{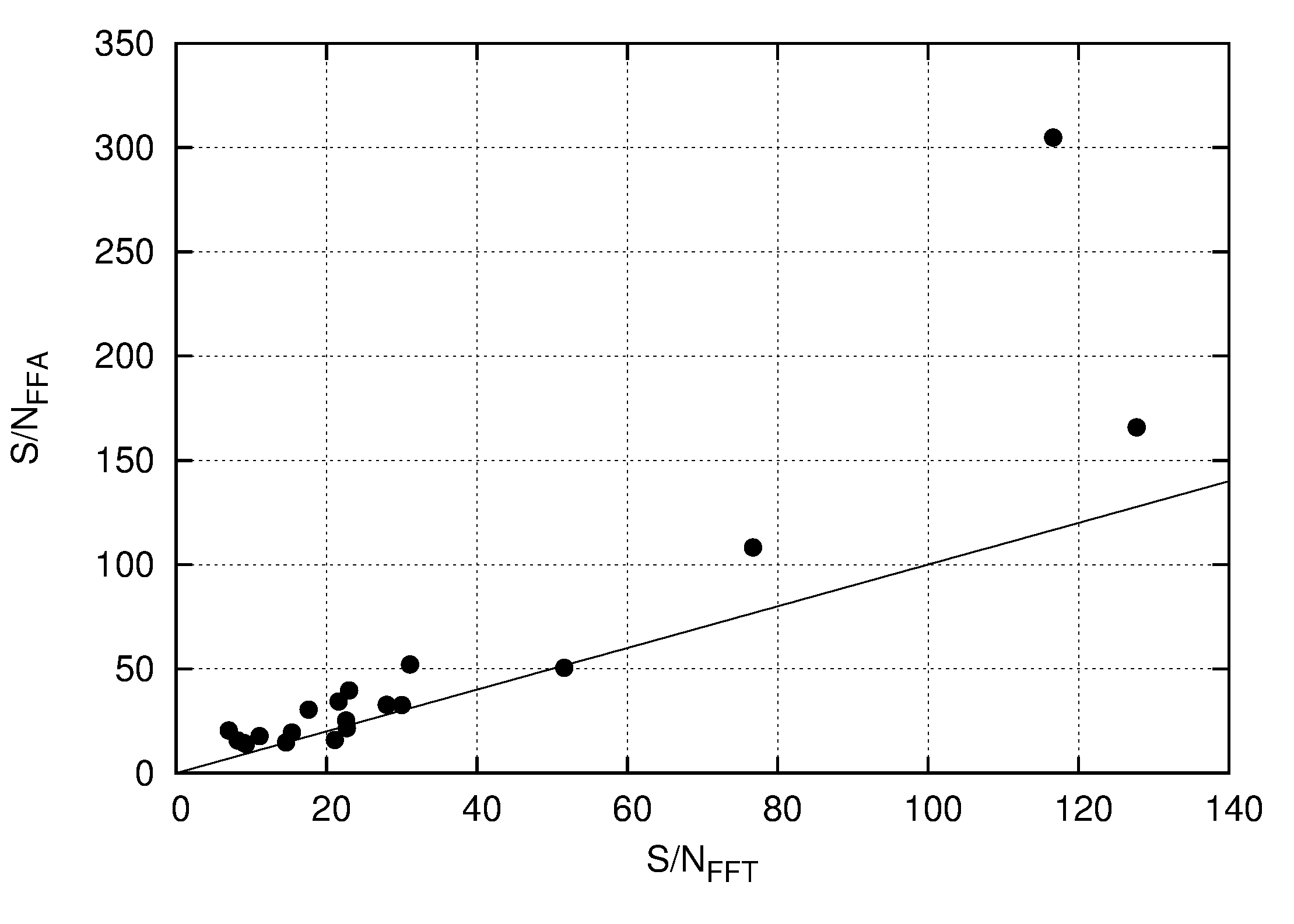}
\caption{Comparison of the sensitivity of FFA Algorithm 1 to the FFT across all 20 pulsars, using the best detections of each technique (including fundamental and harmonic detections). The solid line indicates a relationship of 1:1.}\label{fig:ffavsfft}
\end{figure}

A comparison of the best detections made by both the FFA and the FFT shows a clear trend towards the FFA typically either matching or exceeding the performance of the FFT, especially in the case of Algorithm 1. This trend, expected for long-period pulsars, can be seen in Figure \ref{fig:ffavsfft}. All but three pulsars were detected with a higher S/N through the use of Algorithm 1, the most notable of which was J1324-6302, which was detected using Algorithm 1 at a S/N approximately $2.9$ times that of its FFT detection. Of the three remaining pulsars which were detected more strongly in the FFT, two (J1817-1938 and J1831-1223) were detected by Algorithm 1 within $5\%$ of their FFT S/N values. This could potentially be a result of the approximations introduced into both Algorithms 1 and 2 as described in Section \ref{subsec:Period} and the resultant losses documented in Figure \ref{fig:M6 and M7 period tests}. The third pulsar (J1622-4950) shows a much more notable loss of sensitivity in its Algorithm 1 detection, reaching only $75\%$ of its FFT detection. As with this pulsar's non-detection by Algorithm 2, the cause would again seem to be the unusually wide pulse shape of J1622-4950.

Algorithm 2 displays a much less favorable response than Algorithm 1, on average only approximately matching the performance of the FFT. Of the 18 pulsars detected using Algorithm 2, only 9 were detected with higher S/N values than their FFT detections. The highest of these was again J1324-6302, which was detected at a S/N approximately $2.1$ times higher than its FFT detection. At the other extreme, Algorithm 2 was only able to detect J1814-1744 with approximately $49\%$ of the S/N of the FFT detection.

\begin{figure}
 \includegraphics[width=\columnwidth]{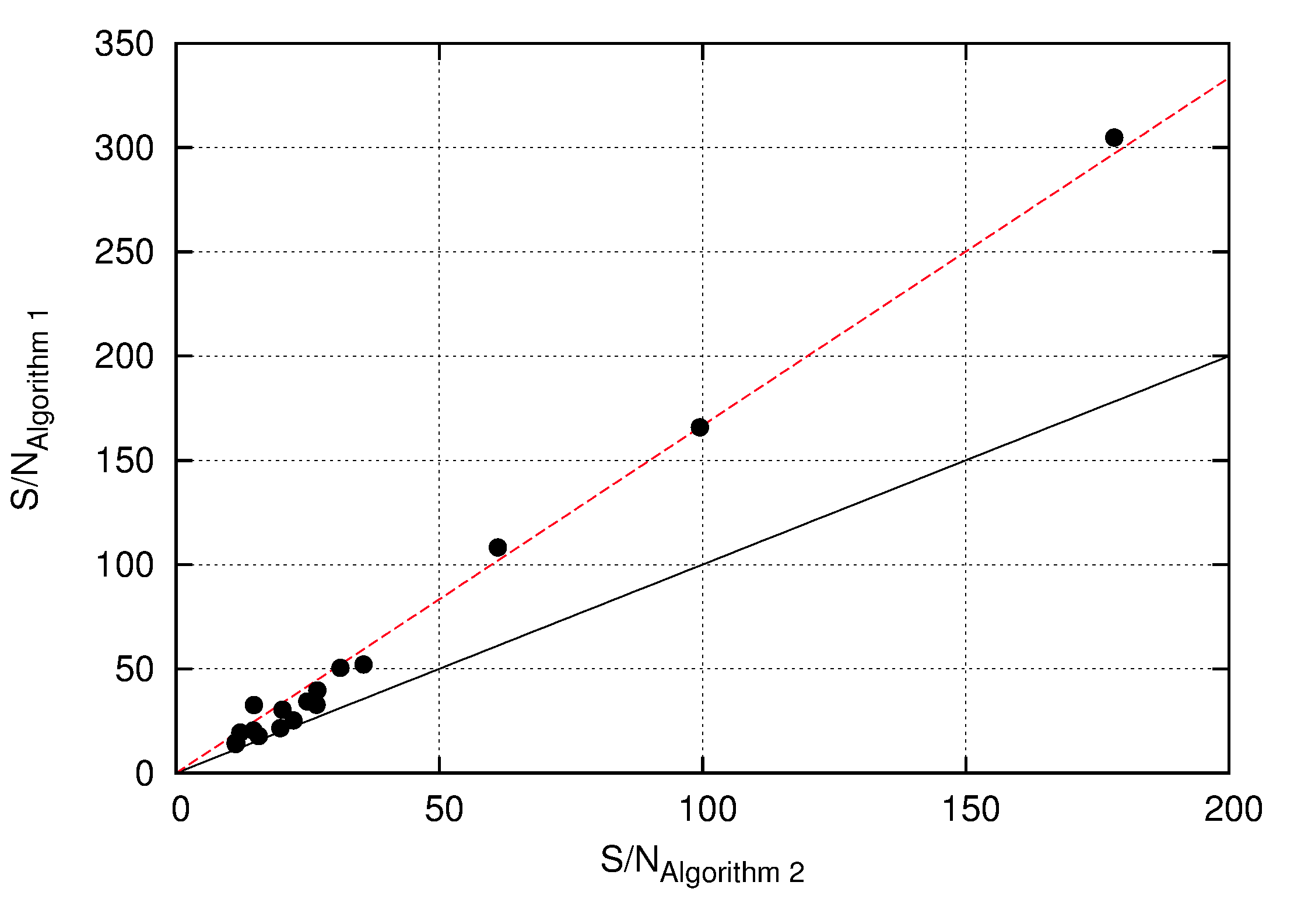}
\caption{Comparison of the sensitivity of Algorithm 1 against Algorithm 2 across the 18 pulsars for which both algorithms were able to make a detection, using the best detections of each technique (including fundamental and harmonic detections). The solid line indicates a relationship of 1:1, while the dashed line indicates a least squares fit to the scatter plot, defined in Equation \protect\ref{eqn: ffa algorithm relation}.}\label{fig:ffavsffa}
\end{figure}

In each of the 18 pulsars detected by both algorithms, Algorithm 1 was able to obtain a higher S/N detection. This relationship is demonstrated in Figure \ref{fig:ffavsffa}, and appears to indicate evidence of an approximately linear trend between the two algorithms. While this relation is only tentative given the limited number of data points, a simple least-squares fit to a linear curve (including a forced intercept at the origin) gives \begin{equation}\label{eqn: ffa algorithm relation} 
               \text{S/N}_{\text{A1}} \simeq 1.668 \times \text{S/N}_{\text{A2}}                                                                                                                                                                                                                                                                                                
                                                                                                                                                                                                                                                                                                              \end{equation} with a coefficient of determination of $R^2 = 0.994$.

\subsubsection{Influence of red noise and other RFI}\label{subsubsec: rfi analysis}

Despite our chosen methods of mitigating both red noise and general RFI, some element of both of these contaminants is likely to still remain in the analysed data. A visual inspection of the folded profiles produced by \texttt{ffancy} indicates only two pulse profiles, J1814-1744 and J1718-3718, which appear to be suffering from visible baseline contamination. An inspection of a manually-folded set of subintegrations reveal this to be caused primarily by persistent RFI which was not removed during the preliminary cleaning procedures. While in the case of J1718-3718 this contamination is likely to have contributed to the pulsar's non-detection by Algorithm 2, both algorithms were able to detect J1814-1744 despite the profile contamination. This is likely due to the higher S/N of the pulse profile being able to stand out against the contaminated baseline. 

\begin{figure}
 \includegraphics[width=\columnwidth]{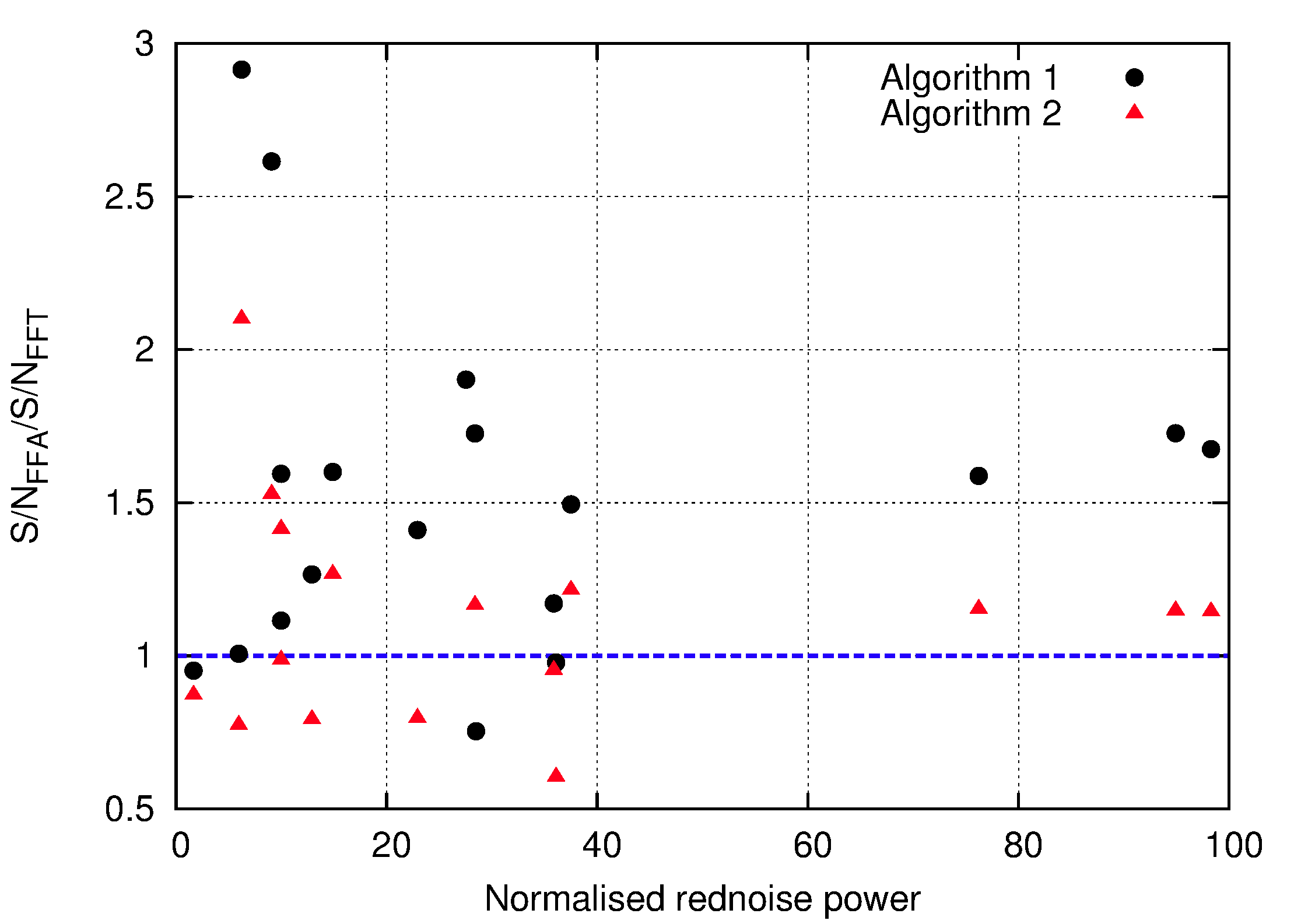}
\caption{Comparison of the ratio of the sensitivity of both FFA Algorithms 1 \& 2 to the FFT against red noise content at each pulsar's fundamental frequency, across all 18 pulsars for which the red noise content could successfully be characterised. The blue line indicates the 1:1 relationship between the FFA and FFT. Above the line, the FFA performs better, while below the line the FFT performs better. The best detections of each technique (including fundamental and harmonic detections) were used.}\label{fig:ffavsred}
\end{figure}

A more analytical presentation of the remaining influence of red noise may potentially be determined by comparing the improvement in detectability of each pulsar against the red noise content of each dataset at the pulsar's fundamental period. This red noise content can be calculated using the parameters listed in Table \ref{tab:HTRU Pulsar Table}, and the results for both Algorithms 1 \& 2 are plotted in Figure \ref{fig:ffavsred}. However, inspection of this figure shows no apparent correlation between the improvement in detectability and red noise content. This could again be a result of the limited size of the dataset employed in this study, as other unaccounted factors may be playing into the detectability of each pulsar. Future comparisons using larger datasets may assist in revealing any underlying trend.

\begin{figure}
 \includegraphics[width=\columnwidth]{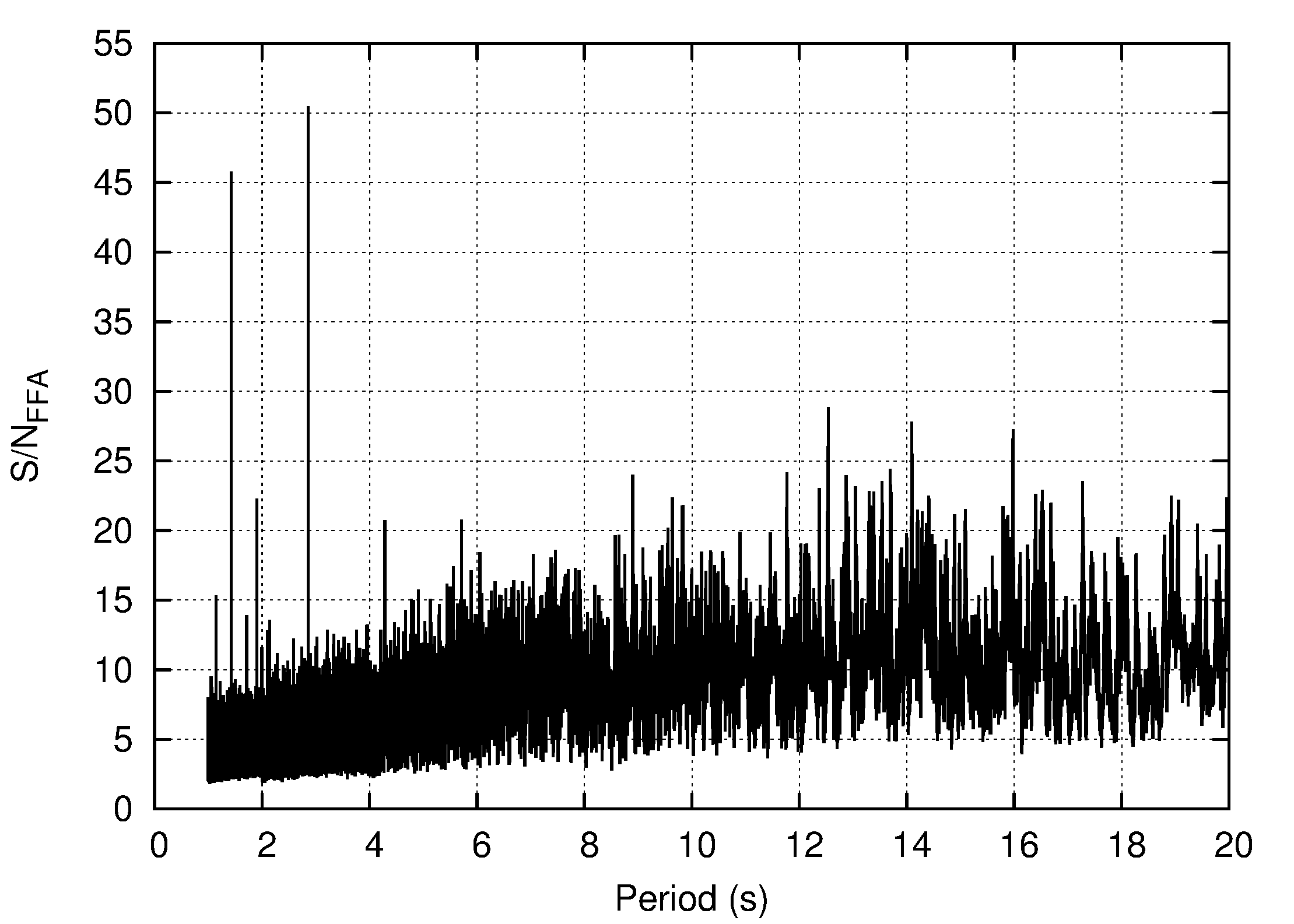}
\caption{Periodogram of a search conducted using Algorithm 1 on pulsar J1831-1223. The pulsar is visible as the strong peak near 2.858 seconds, along with its half harmonic.}\label{fig:pgram_red}
\end{figure}

The persistence of the influence of red noise even in the FFA is also evident upon an examination of the periodograms resulting from each FFA execution. This is particularly the case with Algorithm 1, as its higher response to wider pulse shapes means that it is likely to be more sensitive (in the absence of a stronger true pulsar signal) to the longer profile baseline variations which typically characterise red noise. This effect is demonstrated in Figure \ref{fig:pgram_red}, which shows the Algorithm 1 periodogram for pulsar J1831-1223. Note that at longer periods, the average value of the periodogram appears to rise. This is caused by the continued persistence of red noise creating a pulsar-like signal and causing an undesired response in the algorithm. However, this effect does not persist in the periodograms of Algorithm 2, due to the reasons already described in relation to its pulsar non-detections.

\subsubsection{Signal significance and false-alarm statistics}\label{subsubsec: false alarm}

Up until this point, we have been considering the response of both the FFT and the FFA only in terms of either the spectral or profile-based S/N of each pulsar or candidate signal. However, it is worth considering the statistics of each of these candidate distributions in order to evaluate the detection significance afforded by each technique, as well as the expected corresponding false alarm rates. When combined with the number of trials conducted or candidates produced as part of a given FFT or FFA search, these will allow for the true significance of a candidate to be determined. The statistics for the FFT of normally-distributed white noise are reasonably well understood, and will not be re-derived here (see \cite{lk05} for a basic overview). However, a more in-depth study of the FFA is required.

\begin{figure}
  \includegraphics[width=\columnwidth]{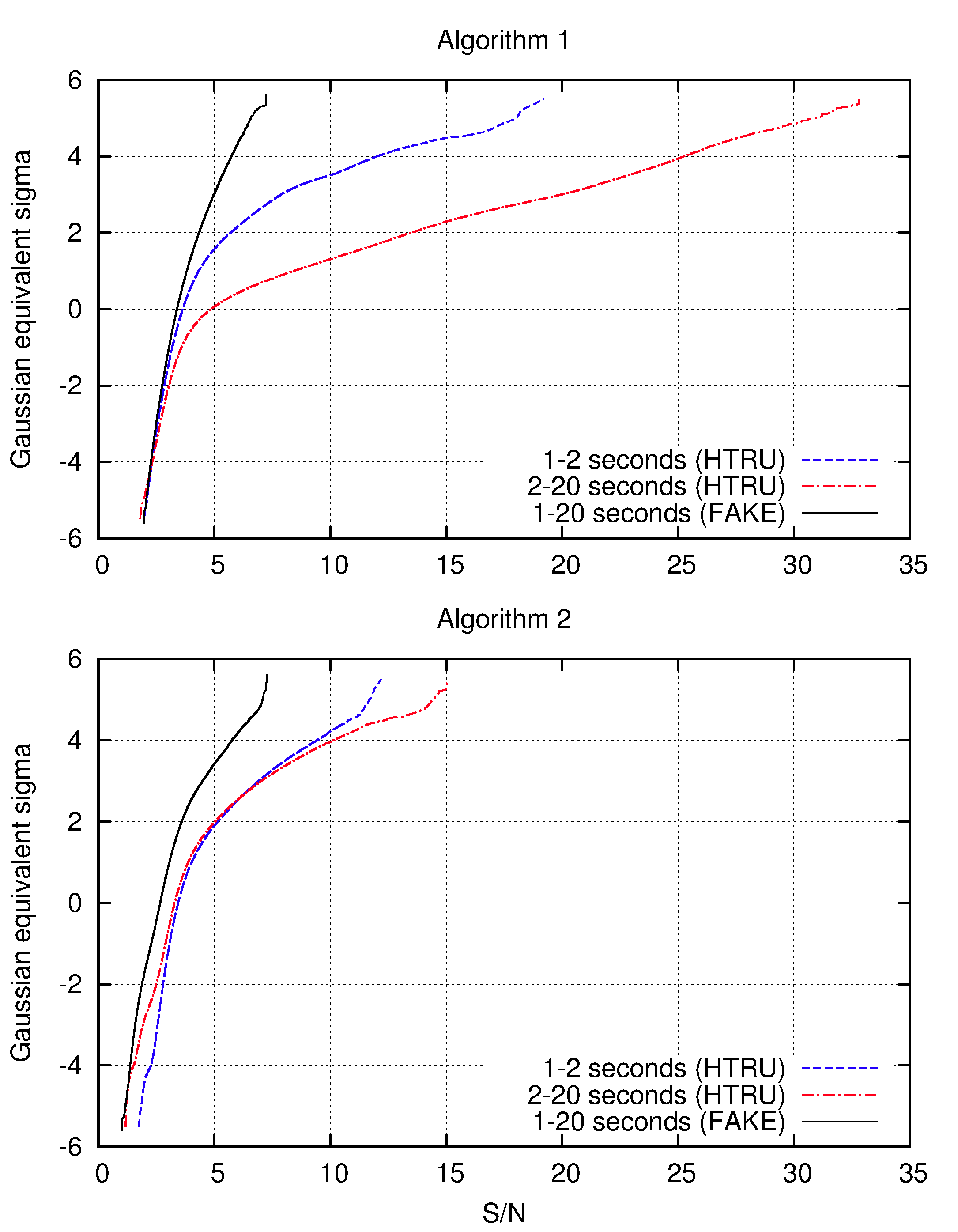}
\caption{A demonstration of the the S/N distribution curves for Algorithms 1 \& 2 map against detection significance, expressed as Gaussian Equivalent Sigma ($\sigma_\text{gauss}$). The FAKE distributions were constructed by deploying \texttt{ffancy} across multiple artificial filterbanks seeded with normally-distributed noise using the \texttt{SIGPROC} program \texttt{fake}. The HTRU distributions were constructed by deploying \texttt{ffancy} over a set of HTRU-S LowLat filterbanks which were devoid of pulsars, dedispersed using a DM of $200 \text{ cm}^\text{-3}\text{pc}$.}\label{fig: distribution histogram}
\end{figure} 

We can evaluate the significance of an FFA algorithm score by studying the distribution of S/N values produced by each algorithm over a large number of trials. For a given operation of the FFA ($N/P_0$ period trials for a given base period $P_0$) conducted across a time series constructed from normally-distributed values, the evaluation of the S/N of a single folded profile can approximately be thought of as the maximum of $n_{bin} \times n_{fil}$ independent normally distributed variables, where $n_{bin}$ is the number of bins in the profile and $n_{fil}$ is the number of applied matched filters. However, while these values remain constant for each incremental period trial in a single FFA execution, these numbers vary significantly across different FFA executions. For example, the specific implementation of the FFA in \texttt{ffancy} causes $n_{bin}$ to vary by as much as a factor of $2$. 

In order to characterise these distributions, 26 normally-distributed filterbank files built to match HTRU specifications were created using \texttt{fake}. These were then integrated in frequency into single-channel time series at a DM of $0 \text{ cm}^\text{-3}\text{pc}$ which were run through \texttt{ffancy} using both Algorithms 1 \& 2 over a search from $1$ to $20\text{ s}$ and an initial downsampling factor of $2^5 = 32$, such that the total number of produced S/N trials was approximately $10^8$. An identical analysis was also performed using 26 representative beams from HTRU-S LowLat which had previously been found to be devoid of pulsars, so as to determine the influence of RFI and other red noise on the distribution. In order to minimise the influence of local RFI (producing a close to best case scenario), these beams were dedispersed at a DM of $200 \text{ cm}^\text{-3}\text{pc}$. Due to the increasing presence of red noise with increasing period, these HTRU distributions were divided into period ranges of $1-2\text{ s}$ and $2-20\text{ s}$, with each range containing approximately the same number of trials ($5\times10^7$). Each value of S/N was converted into a Gaussian Equivalent Sigma ($\sigma_\text{gauss}$) by mapping its probability to a normal quantile function. The resulting distributions can be seen in Figure \ref{fig: distribution histogram}.

While the distributions of the two Algorithms are largely similar in their response to the white noise distributions generated with \texttt{fake}, significant differences can be observed in their response to the real HTRU data and in their response to longer periods and corresponding red noise. As encountered previously, Algorithm 2 appears to be far more robust in its sensitivity to red noise, with its $2-20\text{ s}$ distribution varying very little from its $1-2\text{ s}$ distribution, with only a small increase in S/N at the highest values of $\sigma_\text{gauss}$ (and a corresponding decrease in S/N at the lowest values of $\sigma_\text{gauss}$). A $5\sigma_\text{gauss}$ detection moves from an S/N of $7.1$ in the \text{fake} case to $11.7$ and to $14.5$ in the $1-2\text{ s}$ and $2-20\text{ s}$ distributions respectively.

In contrast, Algorithm 1 shows a much more pronounced response to the transition to real HTRU data and to the increasing presence of red noise. This is likely a reflection of the same response seen in Figure \ref{fig:pgram_red}, once again a result of Algorithm 1's more favorable response to wider pulse profiles. While the responses below $\sigma_\text{gauss}=0$ are very similar, a $5\sigma_\text{gauss}$ detection moves from an S/N of $6.6$ in the \text{fake} case to $18.0$ and to $30.9$ in the $1-2\text{ s}$ and $2-20\text{ s}$ distributions respectively, indicating that at longer periods, pulsar signals may suffer from a reduced apparent significance due to the presence of spurious detections caused by red noise and other interference. 

Unfortunately, due to the lack of an appropriate analytic model of red noise which can be applied generally to multiple telescope configurations, it is presently impractical to develop an analytic expression for deriving these false alarm statistics for general observational data. At this stage, generation of statistical distributions such as the one presented in Figure \ref{fig: distribution histogram}, using data devoid of strong pulsar signals, remains the best solution for producing observation-specific false alarm and significance thresholds.

\section{Discussion}\label{sec:Discussion} 

It is clear that from a theoretical standpoint, the FFA presents many clear advantages over the FFT in the search for long period pulsars. In the case of clean, white data, the two algorithms  which have been presented for consideration in this paper demonstrate a robust ability to detect pulsars typical of the long-period population, over and above the performance of the FFT. However, while the real-world trials demonstrate that this outperformance persists after the transition to real, noisy datasets, there remains considerable room for improvement. 

One area which deserves particular scrutiny is improving our ability to discriminate against and remove forms of interference, including red noise. As noted in our methods, our analysis involved the use of thresholded time and frequency zapping as well as multibeam RFI excision. In addition, we have applied a dynamic red noise removal system through the use of a series of time domain median-filters. While these techniques are well tested and show demonstrable improvement in the resulting signal quality, our analysis shows that they are not entirely successful in the removal of interference.

A particular shortcoming of \texttt{ffancy} (and the FFA in general) in relation to RFI is its inability to discriminate between a signal which is more or less continuous throughout the observation, and a bright, sporadic burst of RFI. A single short instance of strong interference could potentially contaminate a large number of folded profiles with an apparent pulsar-like signal. As the FFA currently only evaluates folded profiles which have been collapsed in time, removing these false positives poses a challenge. One immediate solution which could be applied to future versions of the FFA is a simple clipping mechanism which would cap the values in the initial time series and thereby mitigate any potential contamination. This feature is already available in many pulsar searching packages and could be easily applied to an FFA-based search.

Another proposed solution is to incorporate an additional ``persistence score'' into the FFA. This technique would involve scoring each bin (either in the original time series or in one of the subsequent folding steps of the FFA, depending on the expected strength of the pulsar being searched for) with either a 1 or 0 based upon whether it exceeded a particular cutoff value. In this manner, the combined persistence scores in the bins of a folded profile should rank a continuous pulsar signal more highly than a single burst of strong RFI folded into noise. These scores could either be used to weight the folded profiles, or be written out to be used in later determining which candidates to consider for further investigation. Such ``persistence'' measures could be easily determined during the subsequent folding and inspection (either by person or machine) of each selected candidate from the original observation after the execution of the FFA, but by the incorporation of this step into the FFA itself, the number of false alarm candidates could potentially be reduced at an earlier stage. Such a technique has already been implemented in other pulsar searching algorithms, such as the Pulsar Evaluation Algorithm for Candidate Extraction \citep[PEACE,][]{lsj+13}.

There is also room for improvement in the profile evaluation algorithms themselves. Both FFA algorithms were designed to favour pulsars with a narrow duty cycle, and although this is typically the case for long-period pulsars, exceptions to this general rule were encountered even among the test set of 20 pulsars chosen for study. The most notable example of this was the magnetar J1622-4950, which was missed by Algorithm 2 and suffered a loss of sensitivity in Algorithm 1 because of its large duty cycle. It is these types of scientifically interesting sources that we would most hope to find with future FFA-based searches, and so future versions of these algorithms may need to be modified to allow for better detection of larger duty cycles. 

However, Algorithm 1's existing sensitivity to larger duty cycles caused it to detect a large number of false positive candidates, typically on the order of hundreds for each searched time series. As noted in Sections \ref{subsubsec: rfi analysis} and \ref{subsubsec: false alarm}, Algorithm 1 is more sensitive to the pulsar-like profile baseline variations induced by red noise at longer periods and it tends to produce candidates with an apparent higher significance. This is the source of the reported false positives, and any increase in its duty cycle response is only likely to increase this number of false positive detections. Therefore, significant improvements in this algorithm may not be possible without the implementation of better red noise mitigation techniques.

In theory, the median-filtering technique used by \texttt{ffancy} should have been capable of filtering out the red noise present in the data. However, this technique is only capable of filtering out those red noise signals with periods longer than that of the pulsar itself. Lower periods cannot be easily removed using this technique without potentially removing the desired pulsar signal, and it is this form of red noise which is likely to be producing the red noise detected by Algorithm 1. The multibeam RFI excision technique we have used on sporadic, ground based interference is also not an applicable technique, as the properties responsible for the generation of red noise (such as thermal variability) will be independent to each beam of the receiver. 

An alternative approach to red noise removal potentially lies in the use of interferometers for pulsar observations. While the same problem of independent receiver noise will remain, by forming beams onto multiple adjacent sky locations using the same telescope information, the red noise component which will be common to all beams may be able to be removed while a pulsar, located in only a single beam, would remain. The investigation of such a technique is of particular interest with the development of new, large scale interferometers such as MeerKAT and the SKA, and may lend these systems an advantage in the search for long period pulsars (see e.g. Stappers, B. and Kramer, M. 2017, submitted).

In addition to the two algorithms presented here, the challenge of evaluating a large number of folded candidate profiles is one which has already been partly addressed through machine learning techniques. These techniques have had a history of successful application to pulsar surveys, including the Parkes Multibeam Pulsar Survey (PMPS) \citep{eatough10} and HTRU \citep{bates12}. In particular, we have followed on from work conducted by \cite{zhu+14}, which involves the processing of large numbers of pulsar candidate diagnostic plots through a Pulsar Image-based Classification System (PICS). PICS consists of an ensemble of classifiers based on different machine learning techniques. The system is trained on how to recognise good candidates using a set of human-selected candidate detection plots, and its use has resulted in the discovery of at least six pulsars in the PALFA survey. As part of a preliminary exploration of the application of this technique to the FFA, a newly discovered pulsar from the HTRU-S LowLat survey (Cameron et al. in prep.) with a period of $\simeq1.7\text{ s}$ and a FFT S/N of $\simeq14$ was processed through a FFA search (from approximately $1$ to $10\text{ s}$) with \texttt{ffancy}. The resulting 1.9 million folded profiles from this search were then passed through PICS. The pipeline was able to detect the pulsar as the top ranked candidate after $36s$ of processing time using a neural net classifier running across 20 processing cores, and $~600s$ using a Support Vector Machine (SVM) classifier running on only a single core (giving a similar per-core processing performance). Room exists for further speed optimisation through the use of further parallel processing. Although of limited scope, this test demonstrates a potential usefulness of this technique in future versions of the FFA. However, FFA-specific training sets would also need to be developed in order for this method to proceed further.

Finally, it should be noted that the FFA technique bears strong similarities to image processing and pattern recognition techniques involving the use of the Radon Transform that have been developed in parallel in recent years \citep[e.g.][]{radon2,radon3,radon1}. These techniques (collectively labeled as the Fast or Discrete Radon Transform) use a similar strategy to the FFA in order to remove redundant computations in the processing of the Radon Transform, transforming an $N\times N$ size image through a series of integrals taken along a complete set of lines through the image. This results in the same order of computational complexity as the FFA for a given period, $O(N\log_2(N))$. The analogy of treating aspects of pulsar searching as an image-processing problem is not new, and further investigation of the work done in this field may produce valuable further insights into extracting further performance gains from the FFA.

In the immediate future, this work will now be turning to focus on improving the performance and efficiency of our FFA search techniques. A second, GPU-based FFA software package known as \texttt{ffaster} has been developed in parallel with the CPU-based \texttt{ffancy}, and will incorporate the results of the tests conducted in this study. \texttt{ffaster} should be able to take full advantage of the parallelisability of the FFA so as to gain a significant increase in performance, and will help to cement the FFA as a practical pulsar searching technique for future large-scale pulsar surveys.

\section{Conclusions}\label{sec:Conclusion} 

In this paper, we have conducted an intensive study of the behaviour of FFA against the behaviour of the FFT in both an ideal white noise regime and in a real-world observational regime using data from the HTRU-S Low Latitude pulsar survey. In the ideal white noise regime, two separate profile evaluation algorithms have been characterised and tested in their response to a wide variety of possible pulse shapes, including variances in pulse height, width and total pulse energy as well as the presence of scattering and multiple pulse components, including interpulses. Algorithm 1, a boxcar matched-filter with Median Absolute Deviation (MAD) normalisation, showed response patterns which were significantly more sensitive to wider pulse shapes (including pulses exhibiting scattering tails, as well as to multiple pulse components and to interpulses) than Algorithm 2, a boxcar matched-filter with an off-pulse window. However, both algorithms demonstrated an ability to significantly exceed the performance of the FFT under similar testing regimes, confirming and building upon the results demonstrated by \cite{kml09}.

Trials on real observational data, coupled with a dynamic dereddening scheme employing a running median filter, showed that over a test set of 20 long period pulsars, the FFA has the ability to match and often exceed the performance of the FFT by a maximum factor of almost 3, despite the presence of red noise and other contamination. Algorithm 1 was demonstrated as being the more sensitive of the two tested algorithms, exceeding the performance of the FFT in all but 3 of the test pulsars, but suffered from a large number of false positive candidates, largely due to the influence of red noise. Algorithm 2 was on average only able to match the performance of the FFT when tested on real data, and was unable to detect two of the 20 pulsars in the test sample.

Room for improvement remains on both of the presented Profile Evaluation Algorithms as well as the noise removal techniques used in partnership with the FFA. Opportunity also remains for additional evaluation techniques to be developed which can be coupled with the FFA, including neural network pattern recognition. However, this paper clearly demonstrates the ability of the FFA to recover the sensitivity lost by the FFT in the long-period pulsar regime, and lays a strong foundation for future research into the behaviour of this technique. The custom FFA software package \texttt{ffancy} produced as part of this work will allow for further investigation and testing of the FFA to be conducted as we move forwards towards a faster, GPU-based implementation of the algorithm for use on large-scale pulsar surveys.

\section*{Acknowledgements}

Observational data used in this paper was made available thanks to the High Time Resolution Universe (HTRU) scientific collaboration. The Parkes Observatory, used in the collection of this data, is part of the Australia Telescope National Facility, which is funded by the Commonwealth of Australia for operation as a National Facility managed by CSIRO. Processing for this paper was largely performed using the \texttt{hercules} computing cluster operated by the Max Planck Computing \& Data Facility (MPCDF). The authors would like to thank Patrick Lazarus for the use of his code in evaluating red noise content, and Vladislav Kondratiev for sharing his expertise on both the FFA and FFT and the MPCDF help desk for their constant assistance. AC acknowledges the support of both the International Max Planck Research School (IMPRS) for Astronomy and Astrophysics at the Universities of Bonn and Cologne and the Bonn-Cologne Graduate School of Physics and Astronomy (BCGS). Both AC and EB gratefully acknowledge the support of the Deutscher Akademischer Austauschdienst (DAAD) and Universities Australia (UA) as part of the Australia-Germany Joint Research Cooperation Scheme under contract no. 57218648.

%%%%%%%%%%%%%%%%%%%%%%%%%%%%%%%%%%%%%%%%%%%%%%%%%%



\bibliographystyle{mnras}
\bibliography{andrew-psrrefs} % if your bibtex file is called example.bib

%\begin{thebibliography}{99}
%\end{thebibliography}

%%%%%%%%%%%%%%%%%%%%%%%%%%%%%%%%%%%%%%%%%%%%%%%%%%

%%%%%%%%%%%%%%%%% APPENDICES %%%%%%%%%%%%%%%%%%%%%

%\appendix



%%%%%%%%%%%%%%%%%%%%%%%%%%%%%%%%%%%%%%%%%%%%%%%%%%

% Don't change these lines
\bsp	% typesetting comment
\label{lastpage}
\end{document}